\documentclass[manuscript]{acmart}

\usepackage{subfigure,setspace,url}

\theoremstyle{acmdefinition}
\newtheorem{assumption}[theorem]{Assumption}
\newcommand*{\QEDA}{\hfill\ensuremath{\square}}
\DeclareMathOperator{\sgn}{sgn}

\begin{document}

\setcopyright{rightsretained}

\title{On Optimal Two-Sided Pricing of Congested Networks}
\author{Xin Wang}
\authornote{X. Wang's email address is yixinxa@mail.ustc.edu.cn.}
\affiliation{
  \institution{University of Science and Technology of China}
  \department{School of Computer Science and Technology}
}
\email{yixinxa@mail.ustc.edu.cn}
\author{Richard T. B. Ma}
\affiliation{
  \institution{National University of Singapore}
  \department{School of Computing}
}
\email{tbma@comp.nus.edu.sg}
\author{Yinlong Xu}
\affiliation{
  \institution{University of Science and Technology of China}
  \department{School of Computer Science and Technology}
}
\email{ylxu@ustc.edu.cn}

\begin{abstract}
Traditionally, Internet Access Providers (APs) only charge end-users for Internet access services; however, to recoup infrastructure costs and increase revenues, some APs have recently adopted two-sided pricing schemes under which both end-users and content providers are charged.
Meanwhile, with the rapid growth of traffic, network congestion could seriously degrade user experiences and influence providers' utility.
To optimize profit and social welfare, APs and regulators need to design appropriate pricing strategies and regulatory policies
that take the effects of network congestion into consideration.
In this paper, we model two-sided networks under which users' traffic demands are influenced by exogenous pricing and endogenous congestion parameters and derive the system congestion under an equilibrium.
We characterize the structures and sensitivities of profit- and welfare-optimal two-sided pricing schemes and reveal that 1) the elasticity of system throughput plays a crucial role in determining the structures of optimal pricing, 2) the changes of optimal pricing under varying AP's capacity and users' congestion sensitivity are largely driven by the type of data traffic, e.g., text or video,
and 3) APs and regulators will be incentivized to shift from one-sided to two-sided pricing when APs' capacities and user demand for video traffic grow.
Our results can help APs design optimal two-sided pricing and guide regulators to legislate desirable policies.
\end{abstract} 


\setcopyright{none}
\settopmatter{printacmref=false, printfolios=false}

\maketitle

\renewcommand{\shortauthors}{Wang et al.}

\section{Introduction}

Internet Access Providers (APs) build massive network platforms by which end-users and Content Providers (CPs) can connect and transmit data to each other.
Traditionally, APs use one-sided pricing schemes and obtain revenues mainly from end-users.
With the growing popularity of data-intensive services, e.g., online video streaming and cloud-based applications, Internet traffic has been growing more than \(50\%\) per annum \cite{craig10internet}, causing serious network congestion, especially during peak hours.
To sustain such rapid traffic growth and enhance user experiences, APs need to upgrade network infrastructures and expand capacities.
However, they feel that the revenues from end-users are insufficient to recoup the corresponding costs.
Consequently, some APs have recently shifted towards two-sided pricing schemes, i.e., they start to impose termination fees on CPs' data traffic in addition to charging the end-users.
For example, Comcast\footnote{ComcastXFINITY, http://www.xfinity.com} and Netflix\footnote{Netflix, https://www.netflix.com} reached a paid peering agreement in 2014 \cite{Wyatt-deal}, under which Comcast provides a direct connection to Netflix and improve its content delivery quality for a fee.
Another example is {\it sponsored data} proposed by AT\&T\footnote{AT\&T Sponsored Data, www.att.com/att/sponsoreddata}, under which CPs are allowed to subsidize end-users the fees induced by their data traffic. 
Since subsidizations indirectly transfer value to the APs, sponsored data is really a two-sided pricing scheme in disguise. Although charging CPs directly or indirectly may increase APs' revenues and thus motivate APs to deploy network capacities and alleviate congestion, it has raised concerns over net neutrality \cite{wu2003network}, whose advocates argue that zero-pricing \cite{lee2009subsidizing} on CPs are needed to protect content innovations of the Internet.
Although the U.S. FCC has recently passed the Open Internet Order\footnote{The U.S. FCC's Open Internet Order, https://www.fcc.gov/document/fcc-releases-open-internet-order} to protect net neutrality, existing two-sided schemes such as paid peering and sponsored data are exempt from the ruling, because these pricing schemes are common practices in the Internet transit context and the FCC is not yet clear about
the policy implications on the utilities of various market participants and social welfare. 

Although prior work \cite{Musacchio2009,Altman2011,njoroge2013investment} have studied the economics of two-sided pricing in network markets, the resulting network congestion and its impacts on the utilities of different parties were often overlooked. 
However, the explosive traffic growth has caused severe congestion in many regional and global networks, which degrades end-users' experiences and reduces their data demand. This will strongly affect the profits of APs and the utilities of end-users and CPs. To optimize individual and social utilities, APs and regulators need to reflect the design of pricing strategies and regulatory policies accordingly. So far, little is known about 1) the optimal two-sided pricing structure in a congested network and its changes under varying system parameters, e.g., the users' congestion sensitivity and the APs' capacities, and 2) potential regulations on two-sided pricing for protecting social welfare from monopolistic providers.
To address these questions, one challenge is to accurately capture the endogenous congestion in the network.
Although the level of congestion is influenced by network throughput, the users' traffic demand and throughput are also influenced by network congestion. It is crucial to capture this endogenous congestion so as to faithfully characterize the impacts of two-sided pricing in congested networks.

In this paper, we propose a novel model of a two-sided congested network built by an AP, which transmits data traffic between end-users and CPs.
We model network congestion as a function of AP's capacity and system throughput, which is also affected by the congestion level.
We capture users' population and traffic demand under pricing and congestion parameters and derive an endogenous system congestion under an equilibrium.
Based on the equilibrium model, we analyze the structures of profit-optimal and welfare-optimal two-sided pricing and their sensitivities under varying system environments, e.g., congestion sensitivity of users and capacity of the AP.
We also compare the two types of optimal pricing and derive regulatory implications.
Our main contributions and findings include the following.

\begin{itemize}

\item We derive the congestion equilibrium of two-sided networks, identify a property of elasticity
(Theorem \ref{theorem:unique-congestion} and \ref{theorem:elasticity}), and study the equilibrium dynamics under varying pricing and system parameters (Proposition \ref{proposition:elasticity} and \ref{proposition:pricing-effect}).

\item We characterize the structures of optimal two-sided pricing (Theorem \ref{theorem:KKT-Lerner} and \ref{theorem:social-welfare}) and show that the profit-optimal pricing equalizes the demand hazard rates on the user and CP sides; however, the welfare-optimal counterpart differentiates them based on the elasticity of throughput and per-unit traffic welfare of both sides.

\item We analyze the sensitivities of optimal two-sided prices under varying capacity of APs (Corollary \ref{corollary:profit capacity} and \ref{corollary:welfare capacity}) and congestion sensitivity of users (Corollary \ref{corollary:profit sensitivity} and \ref{corollary:welfare sensitivity}). The results imply that when network traffic is mainly for online video, APs would increase two-sided prices under expanded capacity, while regulators may want to tighten the price regulation on the side of higher market power. However, when network traffic is mostly for text content, they should take the opposite operations.

\item We compare two-sided pricing with the traditional one-sided counterpart. We find that with the growing capacities of APs and demand for video traffic, APs and regulators will have strong incentives to shift from one-sided to two-sided pricing because the benefits of increased profits and social welfare will continue to grow.
\end{itemize}

We believe that our model and analysis could help APs design two-sided pricing schemes in congested networks and guide regulatory authorities to legislate desirable policies.

\section{Related Work}

Several works have studied two-sided pricing in the Internet markets.
Njoroge et al. \cite{njoroge2013investment} showed that two-sided pricing could help APs extract higher profits and maintain higher investment levels than one-sided pricing.
Altman et al. \cite{Altman2011} analyzed the impacts of CP's revenue models, either subscription or advertisement, on AP's pricing strategies.
Choi and Kim \cite{pil2010net} found that expanding capacity will decrease the CP-side price.
In \cite{Musacchio2009}, Musacchio et al. concluded that two-sided pricing is more favorable in terms of social welfare than one-sided pricing when the ratio between CPs' advertising rates and user price sensitivity are extreme.
All of the above works do not consider the impact of network congestion on user throughput, which strongly influences the AP's pricing strategy. In this paper, we characterize the interactions among network congestion, throughput, and price, based on which we analyze both the profit-optimal and welfare-optimal pricing.

Whether APs should be allowed to charge CPs for their content traffic has been a focus of debate on net neutrality \cite{wu2003network}.
To sidestep this debate and extract revenue from CPs, some APs, e.g., AT\&T, have recently provided sponsored data plans, which allow CPs to partially or fully subsidize users the fees induced by their data traffic for increasing market share.
Since the sponsorship offers a way for CPs to subsidize their users and indirectly transfer value to APs, it could be seen as a variant of the two-sided pricing model.
Some work \cite{andrews2013economic,ma2014subsidization,zhang2015sponsored} have studied this variant and showed that it benefits both CPs and APs.
Because APs charge users different prices for content traffic from different CPs, the sponsored data plan is also considered as a type of price discrimination in disguise and has raised concerns from the FCC who says that they will be monitoring and prepared to intervene if necessary \cite{sponsorfcc}.
In this paper, rather than pursuing the price differentiation, we focus on the two-sided pricing under which users are charged entirely based on their traffic volumes.

From the perspective of modeling and analysis, our model extends Rochet and Tirole \cite{Rochet2003}, in which the two-sided network platforms do not incur congestion.
To capture the endogenous network congestion and its effect in Internet markets, we introduce a system congestion under a market equilibrium and use a gain function to model the decline degrees of network throughput under different congestion levels.
Besides, we also analyze the sensitivities of two-sided pricing under the varying network environments, which are instructive and meaningful for APs and regulators to adjust pricing schemes and regulatory policies with the evolution of the Internet.
Ma \cite{richard2014pay} and Chander and Leruth \cite{chander1989optimal} also consider the service markets with congestion externalities. Ma \cite{richard2014pay} analyzed the pay-as-you-go pricing and competition among multiple ISPs. Chander and Leruth \cite{chander1989optimal} studied the quality differentiation strategy of a monopoly provider. 
Both of them focused on the one-sided markets while we consider the more general two-sided markets.

\section{Macroscopic Network Model}\label{sec:model}

We consider a two-sided network platform built by an AP which transmits data traffic between users and CPs.
Unlike physical commodities, the quality of network service is intricately influenced by a negative network effect:
higher traffic will induce a more congested network with worse performance.
To characterize the congestion, we start with a macroscopic model that captures the physical and economic dynamics among the AP, users, and CPs in this section.

\subsection{Basic Terms and Definitions}
As a preliminary, we first introduce two basic economic and statistical terms that will be used in our model.
\begin{definition}[Elasticity]\label{def:elasticity}
For two variables \(x\) and \(y\), the elasticity of $y$ with respect to $x$, or $x$-elasticity of $y$, is defined by
$\displaystyle\epsilon_x^y \triangleq \Big|\frac{x}{y}\frac{\partial y}{\partial x}\Big|$.
\end{definition}

In economics, elasticity captures the responsiveness of a variable $y$ to a change in another variable $x$. Specifically, it can be equivalently expressed as $\epsilon_x^y = |({\partial y}/y)/({\partial x}/x)|$ and interpreted as the absolute value of the percentage change in $y$ (numerator) in response to the percentage change in $x$ (denominator). Intuitively, when the value $\epsilon_x^y$ is higher, \(y\) responds to the change of \(x\) more strongly and we say that \(y\) is more elastic to \(x\).

\begin{definition}[Hazard Rate]\label{def:hazard-rate}
For a function \(y(x)\), the hazard rate of $y$ with respect to $x$ is defined by $\displaystyle \tilde y^x \triangleq -\frac{1}{y}\frac{\partial y}{\partial x}$.
\end{definition}

In statistics, hazard rate is used to measure the rate of decrease in the function \(y\) with respect to the variable \(x\). In particular, it can be expressed as \(y^x = -(\partial y/y)/\partial x\) and interpreted as the proportion of $y$ that is reduced due to a marginal change of $x$. Note that the function \(y\) can have different values of its hazard rate at different starting points of the variable \(x\).

\subsection{Baseline Physical Model}

We denote a metric of congestion, e.g., delay or loss rate, by $\phi$ to model the congestion level of the AP's network.
We denote the AP's user population by \(m\) and the users' average desirable throughput by \(n\), i.e., the maximum amount of data rate consumed under a congestion-free network with \(\phi = 0\).
When network congestion exists, i.e., \(\phi > 0\), the users' desirable throughput might not be fulfilled; and therefore, their achievable throughput is lower than their desirable throughput. We define the users' average achievable throughput under a congestion level \(\phi\) by \(l(\phi) \triangleq n\rho(\phi)\), i.e., the desirable throughput \(n\) multiplied by a gain factor \(\rho(\phi)\).

\begin{assumption}\label{ass:gain}
\(\rho(\phi)\colon \mathbb{R}_+ \mapsto (0,1]\) is a continuously differentiable, decreasing function of \(\phi\). It has an upper bound \(\rho(0) = 1\) and satisfies that \(\displaystyle\lim_{\phi\rightarrow +\infty} \rho(\phi) = 0\).
\end{assumption}

Assumption \ref{ass:gain} states that the {\em throughput gain} or simply the {\em gain} 
decreases monotonically when the network congestion \(\phi\) deteriorates. In particular, the gain and the users' achievable throughput reach the maximum under no congestion.

In practice, users' throughput is usually a mixture of multiple types of traffic flows. Based on the characteristics of applications and protocols, different types of traffic throughput may have dissimilar responses to network congestion.
For instance, inelastic traffic such as online video streaming cannot tolerate high delays and loss rates, and therefore its throughput gain declines sharply with the deterioration of congestion; however, the throughput gain of elastic traffic \cite{scott95fundamental} such as e-mail does not respond to congestion drastically.
Notice that the terms {\em inelastic traffic} and {\em elastic traffic} in networking refer to the traffic whose throughput gains are sensitive and insensitive to congestion, respectively; however, based on the classic definition of elasticity in economics (i.e., Definition \ref{def:elasticity}), traffic throughput has higher congestion elasticity if it is more sensitive to congestion, and therefore the throughput gain of inelastic traffic is more elastic to congestion than that of elastic traffic.

In this paper, we adopt the elasticity defined in economics to characterize how the throughput gain responds to congestion.
By Definition \ref{def:elasticity}, the elasticity $\epsilon_{\phi}^{\rho}$ characterizes the percentage decrease in the gain $\rho$ in response to the percentage increase in the congestion $\phi$.
Based on this characterization, different forms of the gain function \(\rho(\phi)\) with different elasticities $\epsilon_{\phi}^{\rho}$ can be used to model responses of different mixtures of traffic types to congestion.
For example, when users' traffic throughput includes more online video or file sharing traffic, gain functions with higher or lower elasticities can be adopted, respectively.

Given a fixed level of congestion \(\phi\), we define the aggregate network throughput by \(\lambda(\phi)\triangleq ml(\phi) = mn\rho(\phi)\), i.e.,
the product of the number of users \(m\) and the users' average achievable throughput \(n\rho(\phi)\) under the congestion level \(\phi\).
On the one hand, under the given level \(\phi\) of congestion, the network accommodates certain throughput \(\lambda(\phi)\);
on the other hand, the network congestion \(\phi\) is also influenced by this throughput $\lambda$. We denote the AP's capacity by \(\mu\) and characterize the induced system congestion as a function \(\phi \triangleq \Phi(\lambda,\mu)\) of the system throughput \(\lambda\) and capacity \(\mu\).

\begin{assumption}\label{ass:congestion_function}
$\Phi(\lambda,\mu):\mathbb{R}^2_+\mapsto \mathbb{R}_+$ is continuously differentiable, increasing in $\lambda$, decreasing in $\mu$.
\end{assumption}

Assumption \ref{ass:congestion_function} characterizes the physics of network congestion: the congestion level is higher when the system accommodates larger throughput or has less capacity, and vice-versa.
Besides, different forms of the congestion function \(\Phi\) can be adopted to capture the congestion metric based on different models of network services. For example, the function \(\Phi(\lambda,\mu) = 1/(\mu-\lambda)\) models the M/M/1 queueing delay for network services and
 \(\Phi(\lambda,\mu) =\lambda/\mu\) captures the {\em capacity sharing} \cite{chau2010viability} nature of general network congestion.

By far, we have described the system by a triple $(m, n, \mu)$. Because the congestion increases with the network throughput, which decreases with the deteriorated congestion, the resulting system congestion is defined under an equilibrium.

\begin{definition}\label{def:congestion}
$\phi$ is an induced equilibrium congestion of a system $(m, n,\mu)$ if and only if it satisfies \(\phi = \Phi\big(\lambda(\phi),\mu\big)\).
\end{definition}

Definition \ref{def:congestion} states that the system congestion $\phi$ should induce the aggregate throughput $\lambda(\phi)$ such that it leads to exactly the same level of congestion $\phi = \Phi(\lambda,\mu)$.

We define the inverse function of $\Phi(\lambda, \mu)$ with respect to $\lambda$ by $\Lambda(\phi, \mu) \triangleq \Phi^{-1}(\phi, \mu)$. $\Lambda(\phi, \mu)$ can be interpreted as the implied amount of throughput that induces a congestion level $\phi$ for a system with a capacity $\mu$. By Assumption \ref{ass:congestion_function}, $\Lambda(\phi, \mu)$ is strictly increasing in both $\phi$ and $\mu$.
To characterize the system congestion, we define a {\em gap} function $g(\phi)$ between the supply and demand of throughput under a fixed level of congestion $\phi$ by
\begin{equation*}\label{equation:gap}
g(\phi) \triangleq \Lambda\left(\phi,\mu\right) - \lambda(\phi).
\end{equation*}

\begin{theorem}[Congestion Equilibrium]\label{theorem:unique-congestion}
For any system $(m, n,\mu)$, $g(\phi)$ is an increasing function of $\phi$.
The system operates at a unique level of equilibrium congestion $\phi$, which solves $g(\phi)=0$. 
\end{theorem}

Theorem \ref{theorem:unique-congestion} characterizes the uniqueness of the system equilibrium congestion $\phi$ under which the throughput supply $\Lambda(\phi,\mu)$ equals the aggregate throughput demand $\lambda(\phi)$.
Based on Theorem \ref{theorem:unique-congestion}, we denote the unique equilibrium congestion and the corresponding aggregate throughput of the system $(m, n, \mu)$ by $\varphi = \varphi(m, n, \mu)$ and $\lambda = \lambda(m, n, \mu)$, respectively.
We define the marginal change of the throughput gap $g$ due to a marginal change in the congestion $\phi$ by
\begin{equation*}
\frac{\partial g}{\partial \phi} \triangleq  \frac{\partial \Lambda(\phi,\mu)}{\partial \phi} - m n \frac{d\rho(\phi)}{d\phi} > 0,\label{equation:dg}
\end{equation*}
where the first (second) term captures the change of throughput in supply (demand).
Next, we characterize the impacts of the user population $m$, users' average desirable throughput $n$ and system capacity $\mu$ on the system congestion $\varphi$ and throughput $\lambda$ as functions of $\partial g/\partial \varphi$ and $\partial \Lambda/\partial \varphi$ as follows.

\begin{proposition}\label{proposition:elasticity}
The user population $m$'s impacts on the system congestion $\varphi$ and throughput $\lambda$ are
\begin{equation*}
\frac{\partial \varphi}{\partial m} = \left(\frac{\partial g}{\partial \varphi}\right)^{-1} \frac{\lambda}{m} > 0 \quad \text{and} \quad \frac{\partial \lambda}{\partial m} = \frac{\partial \Lambda}{\partial \varphi} \frac{\partial \varphi}{\partial m} > 0.
\end{equation*}
The desirable throughput $n$'s impacts on $\varphi$ and $\lambda$ are
\begin{equation*}
\frac{\partial \varphi}{\partial n} = \left(\frac{\partial g}{\partial \varphi}\right)^{-1} \frac{\lambda}{n} > 0 \quad \text{and} \quad \frac{\partial \lambda}{\partial n} = \frac{\partial \Lambda}{\partial \varphi} \frac{\partial \varphi}{\partial n} > 0.
\end{equation*}
The system capacity $\mu$'s impacts on $\varphi$ and $\lambda$ are
\begin{equation*}
\frac{\partial \varphi}{\partial \mu} = -\frac{\partial \Lambda}{\partial \mu}\left(\frac{\partial g}{\partial \varphi}\right)^{-1} < 0 \quad \text{and} \quad \frac{\partial \lambda}{\partial \mu} = mn\frac{d \rho}{d \varphi}\frac{\partial \varphi}{\partial \mu} > 0.
\end{equation*}
\end{proposition}

Proposition \ref{proposition:elasticity} derives the impacts of $m$, $n$ and $\mu$ on the induced system congestion $\varphi$ and throughput $\lambda$. It states that 1) if the user population $m$ or the desirable throughput $n$ increases, the system congestion and throughput will increase, and 2) if the AP extends its capacity $\mu$, the system congestion will decrease and the system throughput will increase.

\begin{theorem}[Elasticity of Throughput]\label{theorem:elasticity}
Under the equilibrium of any system $(m,n,\mu)$, the elasticities of the system throughput $\lambda$ with respect to the user population \(m\) and the desirable throughput \(n\) are equal and both satisfy \[\epsilon^{\lambda}_{m} = \epsilon^{\lambda}_{n} = \left(1+\frac{|{\partial \lambda}/{\partial \varphi}|}{{\partial \Lambda}/{\partial \varphi}}\right)^{-1}  \in \left(0,1\right].\] 
\end{theorem}

Theorem \ref{theorem:elasticity} shows that under an equilibrium of any physical system $(m,n,\mu)$, the marginal impact of user population on the system throughput $\epsilon_m^{\lambda}$ equals that of the user's desirable throughput on the system throughput $\epsilon_{n}^{\lambda}$. Fundamentally, this result is due to the product form of network throughput $\lambda = m n \rho(\varphi)$ that is symmetric in $m$ and $n$, although both quantities have very different physical natures. Thus, we define this elasticity of system throughput by
\begin{equation}\label{equation:system elasticity}
\epsilon^\lambda \triangleq  \left(1+\frac{|{\partial \lambda}/{\partial \varphi}|}{{\partial \Lambda}/{\partial \varphi}}\right)^{-1} =  \left(1+mn\frac{\left|{d \rho}/{d \varphi}\right|}{{\partial \Lambda}/{\partial \varphi}}\right)^{-1},
\end{equation}
where ${|\partial \lambda}/{\partial \varphi}|$ and ${\partial \Lambda}/{\partial \varphi}$ measure the marginal decrease and increase in the throughput demand and supply with respect to congestion, respectively.
This $\epsilon^\lambda$ can be interpreted as a metric of {\em relative congestion elasticity of throughput demand}: when the traffic demand is elastic, i.e., $|{\partial \lambda}/{\partial \varphi}|$ is large, and the throughput supply is inelastic, i.e., ${\partial \Lambda}/{\partial \varphi}$ is small, $\epsilon^{\lambda} \rightarrow 0$ and the system can accommodate higher throughput mostly due to the elasticity of the demand; otherwise, $\epsilon^{\lambda} \rightarrow 1$ and the elasticity of the supply plays a bigger role in adopting higher throughput.
Notice that our model generalizes that of Rochet and Tirole \cite{Rochet2003}, under which congestion does not exist and the throughput supply can be regarded to be infinitely elastic, i.e., ${\partial \Lambda}/{\partial \varphi}=+\infty$, and therefore, the elasticity of throughput alway satisfies $\epsilon^\lambda=1$.

\subsection{Two-Sided Pricing Model}

We consider usage-based pricing schemes \cite{hande2010pricing} imposed on users and CPs, which are adopted by most wireless APs, e.g., T-Mobile\footnote{T-Mobile, http://www.t-mobile.com} and AT\&T, and some wired APs, e.g., Comcast \cite{comcastusage}.
In our two-sided pricing model, we assume that the AP charges prices of $p$ and $q$ per-unit data traffic to users and CPs, respectively.

On the user side, we model each user by her value \(v_u\) of per-unit traffic and denote the number of the users of value \(v_u\) by \(f_u(v_u)\), which can be regarded as a value density function of users. We assume that a user subscribes to the AP's access service if and only if she can obtain a positive utility, i.e., her value \(v_u\) of per-unit traffic is higher than the price \(p\). As a result, the population \(m\) of active users of the AP is a function of \(p\), defined by
\begin{equation}\label{equation:m}
m(p) \triangleq \int_p^{+\infty} f_u(v_u)dv_u.
\end{equation}

On the CP side, we model each CP by a tuple \((u_c,v_c)\). \(u_c\) is the average desirable throughput of end-users for the CP's content. \(v_c\) is the CP's per-unit traffic value, which models the CP's profit obtained by charging its customers or advertisers. 
Similar to the user side, we denote the density of the CPs of characteristic \((u_c,v_c)\) by \(f_c(u_c,v_c)\) and assume that a CP uses the AP's access service if and only if it can obtain a positive utility, i.e., its value \(v_c\) of per-unit traffic is higher than the price \(q\). Consequently, the average desirable throughput $n$ of end-users for all active CPs is a function of \(q\), defined by
\begin{equation}\label{equation:n}
n(q) \triangleq \int_q^{+\infty}\!\!\!\!\int_0^{+\infty}u_cf_c(u_c,v_c)du_cdv_c.
\end{equation}

From Equation (\ref{equation:m}) and (\ref{equation:n}), we know that the population $m(p)$ of users decreases with the user-side price \(p\), and the average desirable throughput $n(q)$ of users decreases with the CP-side price \(q\).
Since the two-sided prices \(p\) and \(q\) impact the network throughput \(\lambda\) via the user population $m$ and the average desirable throughput $n$, respectively, \(m(p)\) and \(n(q)\) can be interpreted as {\em demand} functions on the user and CP sides. Intuitively, the higher price \(p\) (\(q\)) results in the smaller demand \(m\) (\(n\)) on the user (CP) side.
Furthermore, by Definition \ref{def:hazard-rate}, we adopt the hazard rate $\tilde m^p$ ($\tilde n^q$) to measure the decreasing rate of the demand $m$ ($n$) with respect to the price $p$ ($q$). In addition, the inverse of the hazard rate $1/\tilde m^p$ ($1/\tilde n^q$) is often regarded as the AP's {\em market power} \cite{weyl2010price} on the user (CP) side, which reflects the AP's ability to profitably raise the price of its network service.
Intuitively, if the AP has higher market power on the user (CP) side, i.e., the demand hazard rate $\tilde m^p$ ($\tilde n^q$) is lower, the AP can set a higher price \(p\) (\(q\)) to optimize its profit, since the demand \(m\) (\(n\)) decreases slower with the price.

Under a two-sided pricing $(p,q)$, the aggregate network throughput can be represented as $\lambda(p,q,\phi) \triangleq m(p) n(q)\rho(\phi)$, i.e., the product of the AP's user population \(m(p)\), the users' average desirable throughput \(n(q)\), and the throughput gain factor \(\rho(\phi)\) under the congestion level \(\phi\). 
Furthermore, we can write the unique equilibrium congestion as $\varphi(p,q,\mu)\triangleq \varphi(m(p),n(q),\mu)$ and define the corresponding system throughput by $\lambda(p,q,\mu) \triangleq \lambda(p,q,\varphi(p,q,\mu))$.
As the prices $p$ and $q$ determine the congestion $\varphi(p,q,\mu)$ and throughput $\lambda(p,q,\mu)$ under any fixed capacity \(\mu\), we investigate their impacts on the congestion and the throughput as follows.
\begin{proposition}\label{proposition:pricing-effect}
For a system with a fixed capacity, the system congestion $\varphi$ and throughput $\lambda$ under two-sided prices $p$ and $q$ satisfy
\begin{align*}
&\frac{\partial \varphi}{\partial p} = -\left(\frac{\partial g}{\partial \varphi}\right)^{-1} \lambda \tilde m^p < 0 \ \ \; \text{and} \quad \frac{\partial \lambda}{\partial p} = \frac{\partial \Lambda}{\partial \varphi} \frac{\partial \varphi}{\partial p}<0;\\
&\frac{\partial \varphi}{\partial q} = -\left(\frac{\partial g}{\partial \varphi}\right)^{-1} \lambda \tilde n^q < 0 \quad \text{and} \quad \frac{\partial \lambda}{\partial q} = \frac{\partial \Lambda}{\partial \varphi} \frac{\partial \varphi}{\partial q}<0.
\end{align*}
Furthermore, the price elasticities of throughput $\lambda$ satisfy
\begin{equation*}
\epsilon^{\lambda}_{p}:\epsilon^{\lambda}_{q} = \epsilon^{m}_{p}:\epsilon^{n}_{q}.
\end{equation*}
\end{proposition}

Proposition \ref{proposition:pricing-effect} explicitly shows the impacts of prices $p$ and $q$ on the system congestion $\varphi$ and throughput $\lambda$ and states that the congestion and throughput will decrease if higher prices are charged. Intuitively, higher prices reduce the demands in terms of $m$ and $n$, which further reduce the equilibrium congestion $\varphi$ by Proposition \ref{proposition:elasticity}. As the system has a fixed capacity, the aggregate throughput $\lambda = \Lambda(\varphi, \mu)$ would decrease consequently.
Proposition \ref{proposition:pricing-effect} also tells that the price elasticity of throughput is proportional to that of the corresponding demand. This again shows that the demands $m$ and $n$ play similar roles on the two sides of the market.

\section{Structure of Optimal Pricing}
In the previous section, we modeled a two-sided network in which the effect of congestion is taken into consideration.
In this section, we further explore the structures of the profit-optimal and welfare-optimal pricing in the network.
In particular, we will show the impact of system congestion on the structures of optimal pricing.

\subsection{Structure of Profit-Optimal Pricing}

We first study the optimal two-sided pricing used by the AP to maximize its profit. We assume that the AP incurs a per-unit traffic cost of $c$, which models the recurring maintenance and utility costs like electricity. We define the AP's profit by $U(p,q,\mu) \triangleq (p+q-c)\lambda(p,q,\mu)$, i.e., the per unit traffic profit $p+q-c$ multiplied by the aggregate throughput \(\lambda\). Under any fixed capacity $\mu$, the AP can maximize its profit by determining the optimal prices that solve the following optimization problem.

\begin{equation*}
\begin{aligned}
& \underset{p, q}{\text{maximize}}
& & U(p,q,\mu)=(p+q-c)\lambda(p,q,\mu).
\end{aligned}
\end{equation*}

Before solving the profit maximization problem, we first characterize the impacts of the AP's capacity and prices on its profit as the following result.

\begin{proposition}\label{proposition:profit}
The impact of the capacity $\mu$ on the profit $U$ is
\begin{align*}
&\dfrac{\partial U}{\partial \mu} = (p + q -c)\dfrac{\partial \Lambda}{\partial \mu}(1-\epsilon^\lambda)> 0,
\end{align*}
and the impacts of the prices $p$ and $q$ on the profit $U$ are
\begin{align*}
\begin{cases}
\dfrac{\partial U}{\partial p} =  \lambda - (p+q-c)\epsilon^\lambda\lambda \tilde m^p;\vspace{0.05in}\\
\dfrac{\partial U}{\partial q} = \lambda - (p+q-c)\epsilon^\lambda\lambda \tilde n^q.
\end{cases}
\end{align*}
\end{proposition}

Proposition \ref{proposition:profit} intuitively shows that the AP's profit increases with its capacity under fixed prices; however, increasing prices might reduce demands, which could either increase or decrease the profit.
Next, we characterize the optimal two-sided prices of the AP that maximize its profit.

\begin{theorem}\label{theorem:KKT-Lerner}
If prices $p$ and $q$ maximize the AP's profit, the following condition must hold:
\begin{equation}\label{equation:KKT-necesary}
\displaystyle {\tilde m^p}   = {\tilde n^q} = \frac{1}{ (p+q-c)\epsilon^\lambda}.
\end{equation}
Furthermore, the total price $p+q$ satisfies
\begin{equation}\label{equation:Lerner-formula}
\frac{p+q-c}{p+q}  =  \frac{1}{\epsilon_p^\lambda + \epsilon_q^\lambda} =  \frac{1}{\epsilon^\lambda (\epsilon_p^m + \epsilon_q^n)}.  
\end{equation}
\end{theorem}

Theorem \ref{theorem:KKT-Lerner} provides necessary conditions for prices to be profit-optimal. Equation (\ref{equation:KKT-necesary}) shows the connections among the optimal prices,
the demand hazard rates on both sides, and the elasticity of system throughput. In particular, the optimal prices will equalize the hazard rates of demands on both sides, which equals the inverse of the product of the profit margin $p+q-c$ and the elasticity of system throughput $\epsilon^\lambda$. This implies that the AP will always balance its market power on both sides of the market so as to maximize its profit.
Notice that the formula of Equation (\ref{equation:KKT-necesary}) also generalizes the result of Rochet and Tirole \cite{Rochet2003} for the structure of the profit-optimal two-sided pricing under endogenously congested networks.
Besides, Equation (\ref{equation:Lerner-formula}) characterizes the relationship between the elasticities and price margins of the profit-maximizing AP, where the total price $p+q$ follows a form of the Lerner index \cite{Lerner34}, i.e., the ratio of profit margin to price equals the inverse of the total elasticities of the system throughput with respect to the prices.

By Equation (\ref{equation:KKT-necesary}), we see that the profit-optimal two-sided prices are related to the elasticity of system throughput $\epsilon^\lambda$, which depends on the level of network congestion by Equation (\ref{equation:system elasticity}). To see more clearly, we illustrate it with an example where the gain function is \(\rho(\phi)=e^{-\phi}\), the congestion function is \(\Phi(\lambda,\mu) = \lambda/\mu\) and the demand functions are \(m(p) = 1-p\) and \(n(q) = (1-q)^2\) (\(p,q\in [0,1]\)). For this example, we can derive the explicit profit-optimal prices based on Theorem \ref{theorem:KKT-Lerner}:
\begin{equation}\label{equation:example profit}
p = \frac{\varphi+c+2}{\varphi+4} \quad \text{and} \quad q = \frac{\varphi+2c}{\varphi+4}
\end{equation}
where \(\varphi\) is the equilibrium congestion of the system.
Equation (\ref{equation:example profit}) shows that the congestion level directly affects the profit-optimal prices. Therefore, the AP should fully take the congestion effect into consideration when designing two-sided pricing schemes.

\subsection{Structure of Welfare-Optimal Pricing}
We next analyze the welfare-optimal pricing structure that maximizes social welfare and contrast it with the profit-optimal counterpart.
On the user side, given any fixed price \(p\), a user of value \(v_u\) obtains the surplus \((v_u-p)\) for per-unit traffic, and therefore the total surplus of all users, when each of them consumes one unit traffic, can be defined by
\begin{align}\label{equation:def_sm}
S_m(p) \triangleq  \int^{+\infty}_p (v_u-p)f_u(v_u)dv_u,
\end{align}
where \(f_u(v_u)\) measures the population of the users of value \(v_u\) for per-unit traffic. Thus the per-user average surplus for per-unit traffic can be defined by
\(s_m(p) \triangleq S_m(p)/m(p)\), where \(m(p)\) is the user population.
Accordingly, we define the total user welfare for the aggregate throughput by
\begin{equation*}
W_m (p, q,\mu) \triangleq s_m(p)\lambda(p,q,\mu) = S_m(p) n(q) \rho\big(\varphi(p,q,\mu)\big),
\end{equation*}
i.e., the users' average per-unit traffic surplus \(s_m(p)\) multiplied by the aggregate traffic throughput \(\lambda(p,q,\mu)\).
Similarly, on the CP side, given any fixed termination fee \(q\), a CP of characteristic \((u_c,v_c)\) generates a surplus \((v_c-q)\) for per-unit traffic and the average desirable throughput per user on this CP is \(u_c\); and therefore, the CP can obtain a surplus of \((v_c-q)u_c\) per user. As a result, the total surplus of all active CPs, when each of them accommodates their users' average desirable throughput, can be defined by
\begin{align*}
S_n(q) \triangleq &\int_q^{+\infty}\!\!\!\!\int_0^{+\infty}(v_c-q)u_cf_c(u_c,v_c)du_cdv_c,
\end{align*}
where \(f_c(u_c,v_c)\) measures the population of the CPs of characteristic \((u_c,v_c)\). Thus, the total surplus of all active CPs for per-unit traffic can be defined by \(s_n(q) \triangleq S_n(q)/n(q)\), where \(n(q)\) is the end-users' average desirable throughput on all active CPs.
Accordingly, we define the total CP welfare generated from all end-users by
\begin{equation*}
W_n (p, q,\mu) \triangleq s_n(q)\lambda(p,q,\mu) = S_n(q) m(p) \rho\big(\varphi(p,q,\mu)\big),
\end{equation*}
i.e., the aggregate per-unit traffic CP surplus \(s_n(q)\) multiplied by the aggregate traffic throughput \(\lambda(p,q,\mu)\).
Notice that because the welfare \(W_m(p,q,\mu)\) and \(W_n(p,q,\mu)\) of the two sides are the system throughput \(\lambda(p,q,\mu)\) multiplied by \(s_m(p)\) and \(s_n(q)\), respectively, \(s_m\) and \(s_n\) can also be interpreted as the {\em per-unit traffic welfare} of the user and CP sides.
Based on the above definitions, social welfare can be denoted as the summation of the end-users' welfare, the CPs' welfare, and the AP's profit, defined by
\[W(p, q,\mu) \triangleq W_m (p, q,\mu)+W_n (p, q,\mu)+ U(p,q,\mu).\]

Because social welfare usually increases as the two-sided prices decrease, it might be maximized at a point where the AP incurs a loss. 
To ensure that the AP does not incur a loss and the welfare-optimal pricing scheme is practically feasible, we consider the framework of Ramsey-Boiteux pricing \cite{ramsey1927contribution} which tries to maximize social welfare, subject to a constraint on the AP's profit.
In our context, we typically constrain the AP's profit to be zero, i.e., \(U=0\), under which the total price \(p+q\) is equal to the traffic cost \(c\).
We formulate the welfare optimization problem as follows.
\begin{align*}
& \underset{p, q}{\text{maximize}} \quad\, W(p, q,\mu) = W_m(p,q,\mu) + W_n(p,q,\mu)\\
& \text{subject to} \quad\ p+q=c.
\end{align*}

Before we solve the welfare maximization problem, we first characterize the impacts of the capacity and prices on social welfare as follows.

\begin{proposition}
The impact of the AP's capacity $\mu$ on social welfare $W$ is
\begin{align*}
&\dfrac{\partial W}{\partial \mu} = W \tilde{\rho}^\varphi \frac{\partial \Lambda}{\partial \mu} \left(\frac{\partial g}{\partial \varphi}\right)^{-1}> 0.
\end{align*}
The impacts of the AP's prices $p$ and $q$ on social welfare $W$ are
\begin{align*}
\begin{cases}
\dfrac{\partial W}{\partial p} =  -\lambda - \tilde{m}^p\big[W_n -W(1-\epsilon^\lambda)\big]; \vspace{0.05in}\\
\dfrac{\partial W}{\partial q} =  -\lambda - \tilde{n}^q \big[W_m -W(1-\epsilon^\lambda)\big].
\end{cases}
\end{align*}
In particular, if \(\tilde{S}_m^p\) and \(\tilde{S}_n^q\) increase with \(p\) and \(q\), respectively, social welfare $W$ decreases with \(p\) and \(q\).
\label{proposition:welfare}
\end{proposition}

Proposition \ref{proposition:welfare} intuitively shows that 1) under fixed prices, social welfare increases with the AP's capacity, and 2) under the commonly assumed monotone conditions \cite{Barlow63} on the hazard rates \(\tilde{S}_m^p\) and \(\tilde{S}_n^q\) of surpluses, social welfare decreases as the two-sided prices increase. It implies that to protect social welfare, regulators should encourage APs to expand capacity and regulate the prices of both sides.
Next, we characterize the optimal two-sided prices which maximize social welfare.

\begin{theorem}\label{theorem:social-welfare}
If prices $p$ and $q$ maximize the social welfare $W(p,q)$, the following condition must hold:
\begin{equation}\label{equation:welfare}
{\tilde{m}^p}: {\tilde{n}^q} = \left(\epsilon^\lambda-1+\frac{s_m}{s_m+s_n}\right) : \left(\epsilon^\lambda-1+\frac{s_n}{s_m+s_n}\right) .
\end{equation}
\end{theorem}

Theorem \ref{theorem:social-welfare} provides a necessary condition for the prices to be welfare-optimal.
Equation (\ref{equation:welfare}) describes the relationship among the elasticity of system throughput, the demand hazard rates and per-unit traffic welfares of the two sides. Compared to the result of Theorem \ref{theorem:KKT-Lerner} where the profit-optimal pricing always equalizes the demand hazard rates on both sides, the welfare-optimal counterpart differentiates them based on the per-unit traffic welfares of the two sides as well as the elasticity of system throughput.
To see this more clearly, we consider a special case of \(\epsilon^\lambda = 1\), i.e., the model of Rochet and Tirole \cite{Rochet2003} where network congestion does not exist. Under this case, Equation (\ref{equation:welfare}) simplifies to the condition in Proposition 2 of \cite{Rochet2003}, i.e.,
\begin{equation}\label{equation:case 1}
\frac{s_m}{\tilde{m}^p} = \frac{s_n}{\tilde{n}^q}.
\end{equation}

Equation (\ref{equation:case 1}) shows that when the congestion does not exist, the demand hazard rate is proportional to the per-unit traffic welfare of the same side under the welfare-optimal pricing. This implies that the welfare-optimal prices result in a higher demand hazard rate, and therefore induce lower market power, on the side of higher welfare. 
When the network becomes mildly congested, the elasticity of system throughput \(\epsilon^\lambda\) will slightly decline below $1$. 
If the AP has higher per-unit traffic welfare on the user side, i.e., \(s_m>s_n\), the right-hand side of Equation (\ref{equation:welfare}) will increase. 
Since Equation (\ref{equation:welfare}) has to hold under the welfare-optimal pricing, the prices need to adjust to balance the equation again, i.e., 
to increase the ratio of demand hazard rates on the left-hand side of Equation (\ref{equation:welfare}). This implies that the market power of the side of high per-unit traffic welfare will decrease, and vice versa. 
In general, we can expect that when the network becomes more congested, the elasticity of throughput decreases, and under the welfare-optimal pricing, the ratio of the demand hazard rates of the sides with higher and lower per-unit traffic welfare will further increase.

Similar to the profit-optimal prices, the welfare-optimal prices are also related with the elasticity of throughput \(\epsilon^\lambda\) by Equation (\ref{equation:welfare}), which depends on the level of network congestion by Equation (\ref{equation:system elasticity}). Under the same example where the gain function is \(\rho(\phi)=e^{-\phi}\), the congestion function is \(\Phi(\lambda,\mu) = \lambda/\mu\) and the demand functions are \(m(p) = 1-p\) and \(n(q) = (1-q)^2\) (\(p,q\in [0,1]\)), the explicit welfare-optimal prices can be derived based on Theorem \ref{theorem:social-welfare}:
\begin{equation}\label{equation:example welfare}
p = \frac{3k(\varphi)+2c-2}{3k(\varphi)+2} \quad \text{and} \quad q = \frac{3(c-1)k(\varphi)+2}{3k(\varphi)+2}
\end{equation}
where \(k(\varphi)\) is the positive solution of \(3k^2 + \varphi k - 4 =0\) and \(\varphi\) is the equilibrium congestion of the system. Equation (\ref{equation:example welfare}) shows that the congestion level also directly affects the welfare-optimal prices. Thus, regulators need to consider the congestion effect adequately when making the regulatory policies on the two-sided pricing.

\textbf{Summary of Implications:} The theoretical results in this section could help APs and regulatory authorities to design two-sided pricing strategies and the corresponding regulatory policies.
First, to optimize their profits, APs should set two-sided prices to equalize the demand hazard rates at both sides, whose optimal value changes with the elasticity of system throughput (by Theorem \ref{theorem:KKT-Lerner}).
Second, to protect social welfare, regulators might want to regulate the prices of both sides (by Proposition \ref{proposition:welfare}).
In particular, the prices should be regulated such that the difference in the demand hazard rates of the two sides will enlarge as the elasticity of system throughput decreases (by Theorem \ref{theorem:social-welfare}).

\section{Sensitivity of Optimal Pricing}

With the rapid development of Internet, characteristics of APs, CPs, and end-users are continuously changing. For example, APs are using new wireless technologies, e.g., 4G or 5G, to expand their capacities, and as the real-time video traffic grows rapidly, users often become more sensitive to network congestion. In this section, we explore the sensitivities of the profit-optimal and welfare-optimal pricing under these varying characteristics of the market participants.

Because Theorem \ref{theorem:KKT-Lerner} and \ref{theorem:social-welfare} in the previous section only provide necessary conditions for the two-sided prices $p$ and $q$ to be profit- and welfare-optimal, they are not sufficient to guarantee the optimality. To analyze how the optimal prices change with varying parameters, we make the following assumption on the hazard rates of the demands $m$ and $n$ so as to guarantee the (local) optimality of the prices $p$ and $q$.

\begin{assumption}\label{ass:hazard rate}
The demand hazard rates \(\tilde{m}^p\) and \(\tilde{n}^q\) are increasing in the prices \(p\) and \(q\), respectively.
\end{assumption}

The monotonicity conditions on the demand hazard rates \(\tilde{m}^p\) and \(\tilde{n}^q\) stated in Assumption \ref{ass:hazard rate} indicate that at a higher price level \(p\) (\(q\)), the proportion of demand \(m\) (\(n\)) reduced due to a marginal increase in price is larger.
These monotone conditions are widely assumed in various contexts in statistics \cite{Barlow63} and economics, e.g., Myerson's optimal auction \cite{myerson81}. In particular, the hazard rate of any concave function satisfies the monotone property, e.g., a concave demand \(m(p)\) implies that the hazard rate \(\tilde{m}^p\) must be increasing in \(p\).
Under Assumption \ref{ass:hazard rate}, we consider a system that has unique profit-optimal and welfare-optimal two-sided prices, denoted by \((p^*,q^*)\) and \((p^\circ,q^\circ)\), respectively.

As the previous section shows that the structure of optimal pricing depends on the elasticity of throughput $\epsilon^\lambda$, we will see in this section that the sensitivities of optimal prices largely depend on the changing direction of $\epsilon^\lambda$ as congestion increases, i.e., the sign of $\partial \epsilon^\lambda/ \partial \varphi$.
This quantity is determined by the type of data traffic:
if the network traffic is mostly video (text), its throughput gain $\rho(\varphi)$ often decreases convexly (concavely) in $\varphi$, as its traffic is quite sensitive (insensitive) to mild congestion. As a result, we will show that $\epsilon^\lambda$ often increases (decreases) when congestion increases.

\subsection{Impact of AP's Capacity}\label{section:5.1}
In this subsection, we first show how the sensitivity of optimal pricing under varying AP's capacity is impacted by the changing direction of $\epsilon^\lambda$ under changing congestion. We then explain why this changing direction is determined by the type of data traffic.
The following two corollaries show the impacts of the AP's capacity \(\mu\) on the optimal prices.

\begin{corollary}\label{corollary:profit capacity}
The derivatives of the profit-optimal prices \(p^*\) and \(q^*\) with respect to the capacity \(\mu\) satisfy that
\begin{align*}
\sgn \Big(\frac{\partial p^*}{\partial \mu}\Big)=\sgn \Big(\frac{\partial q^*}{\partial \mu}\Big) = \sgn \Big(\frac{\partial\epsilon^\lambda}{\partial\varphi}\Big).
\end{align*}
Furthermore, their ratio satisfies that
\begin{equation}\label{equation:proportion capacity}
\frac{\partial p^*}{\partial \mu}: \frac{\partial q^*}{\partial \mu} = \frac{d \tilde n^q}{d q}:\frac{d \tilde m^p}{d p}.
\end{equation}
\end{corollary}

Corollary \ref{corollary:profit capacity} shows that the signs of the marginal profit-optimal prices ${\partial p^*}/{\partial \mu}$ and ${\partial q^*}/{\partial \mu}$ (with respect to capacity) are the same as that of the marginal elasticity of system throughput ${\partial\epsilon^\lambda}/{\partial\varphi}$ (with respect to congestion).
This result implies that if the elasticity of system throughput increases with deteriorated congestion, the AP's profit-optimal prices will increase with its capacity, and vice-versa.
In general, the system congestion \(\varphi\) will be alleviated with the expanded capacity \(\mu\) by Proposition \ref{proposition:elasticity}. If the elasticity of throughput decreases (increases) with alleviated congestion, i.e., \(\partial\epsilon^\lambda/\partial\varphi>0\) (\(\partial\epsilon^\lambda/\partial\varphi<0\)), the marginal increase of throughput demand $|\partial \lambda/\partial\varphi|$ increases faster (slower) than that of throughput supply $\partial \Lambda/\partial\varphi$ by Equation (\ref{equation:system elasticity}). As a result, the profit-optimal prices will increase (decrease) as the basic economics principle of demand and supply implies.

Furthermore, Equation (\ref{equation:proportion capacity}) shows that the marginal profit-optimal prices (with respect to capacity) are proportional to the marginal demand hazard rate (with respect to price) on the opposite sides.
Because the profit-optimal prices always equalize the demand hazard rates of both sides by Theorem \ref{theorem:KKT-Lerner}, i.e., \(\tilde{m}^p(p^*) = \tilde{n}^q(q^*)\), the changes in demand hazard rates with respect to capacity \(\mu\) should also be the same at both sides. Mathematically, this balanced marginal effect can be used to deduce Equation (\ref{equation:proportion capacity}) and expressed as
\begin{align*}
\displaystyle\frac{\partial p^*}{\partial \mu}\frac{d \tilde m^p}{d p} = \frac{\partial \tilde m^p}{\partial \mu}= \frac{\partial \tilde n^q}{\partial \mu}=\frac{\partial q^*}{\partial \mu}\frac{d \tilde n^q}{d q}.
\end{align*}

\begin{corollary}\label{corollary:welfare capacity}
The derivatives of the welfare-optimal prices \(p^\circ\) and \(q^\circ\) with respect to the capacity \(\mu\) satisfy that
\begin{align*}
\sgn \Big(\frac{\partial p^\circ}{\partial \mu}\Big) = -\sgn \Big(\frac{\partial q^\circ}{\partial \mu}\Big) = \sgn \left(\tilde m^p - \tilde n^q\right)\cdot\sgn\left(\frac{\partial \epsilon^\lambda}{\partial \varphi}\right).
\end{align*}
\end{corollary}

Corollary \ref{corollary:welfare capacity} shows that the marginal prices \(\partial p^\circ /\partial \mu\) and \(\partial q^\circ /\partial \mu\) with respect to capacity have opposite signs. This is due to the constraint of fixed total price of the two sides, which further implies that the values of \(\partial p^\circ/\partial \mu\) and \(\partial q^\circ/\partial \mu\) are always opposite.
Corollary \ref{corollary:welfare capacity} also shows that the sign of marginal price of the side of higher (lower) demand hazard rate will be the same as (opposite to) that of the marginal elasticity of throughput.
This implies that if the elasticity of throughput increases (decreases) with congestion, the welfare-optimal price of the side whose demand hazard rate is higher would increase (decrease) with the capacity.
As explained earlier, if the elasticity of throughput increases with expanded capacity and alleviated congestion, i.e., \(\partial \epsilon^\lambda/\partial \varphi<0\), the marginal increase of throughput demand $|\partial \lambda /\partial \varphi|$ increases slower than that of throughput supply $\partial \Lambda /\partial \varphi$. As a result, the network would be under-utilized if the two-sided prices are unchanged, and therefore under fixed total price, the price of the side of higher demand hazard rate should be reduced to increase system throughput so as to maximize social welfare. Similarly, if \(\partial \epsilon^\lambda/\partial \varphi>0\), the network would be overloaded and the price of the side of higher demand hazard rate should be increased to reduce the throughput.

By comparing the results of Corollary \ref{corollary:profit capacity} and \ref{corollary:welfare capacity}, we see that if the elasticity of throughput $\epsilon^\lambda$ increases (decreases) with congestion $\varphi$, the price of the side of lower demand hazard rate will decrease (increase) with capacity under the welfare-optimal pricing but will increase (decrease) with capacity under the profit-optimal counterpart.
This suggests that, to protect social welfare, if $\epsilon^\lambda$ increases (decreases) with $\varphi$, regulators might want to tighten (relax) the price regulation on the side of higher AP market power, i.e., the side of lower demand hazard rate, when APs expand capacities.

From Corollary \ref{corollary:profit capacity} and \ref{corollary:welfare capacity}, we have seen that under expanded capacity, the sensitivities of both profit- and welfare-optimal prices really depend on the changing direction of $\epsilon^\lambda$ with deteriorated congestion, i.e., the sign of \(\partial \epsilon^\lambda/\partial \varphi\). Next, we explain why it is determined by the type of data traffic. From Equation (\ref{equation:system elasticity}), the elasticity of throughput $\epsilon^\lambda$ is a function of the marginal gain \(|\partial \rho/\partial \varphi|\) and supply \(\partial \Lambda/\partial \varphi\) of throughput.
The former and the latter are determined by the type of data traffic, e.g., text or video, and the congestion model of network service, e.g., capacity sharing \cite{chau2010viability} or M/M/1 queue, respectively.
Since the congestion model is usually fixed, while the data type changes rapidly with the emerging new applications and contents, we focus on the impact of data traffic type.
For example, we consider the congestion function \(\Phi(\lambda,\mu) = \lambda/\mu\) that captures the capacity sharing nature of network services.
As mentioned at the beginning of this section, when the data traffic is mostly for online video (text content), the throughput gain $\rho(\varphi)$ often decreases convexly (concavely) in congestion $\varphi$ and thus is more (less) elastic as the congestion is milder. Consequently, the congestion elasticity of gain \(\epsilon^\rho_\varphi\) usually decreases (increases) with congestion. The next proposition builds the relationship between \(\epsilon^\rho_\varphi\) and the elasticity of system throughput \(\epsilon^\lambda\).

\begin{proposition}\label{proposition:elasticity relation}
If the network congestion meets the form $\Phi(\lambda,\mu)=\lambda/\mu$, the elasticity of system throughput satisfies that
\begin{equation*}\label{equation:elasticity relation}
\epsilon^\lambda = \frac{1}{1+\epsilon^\rho_\varphi}\in (0,1].
\end{equation*}
\end{proposition}

Proposition \ref{proposition:elasticity relation} implies that under capacity sharing, the elasticity of throughput \(\epsilon^\lambda\) will increase as the congestion elasticity of gain \(\epsilon^\rho_\varphi\) decreases. Furthermore, if the network traffic is mostly for text content (online video), the congestion elasticity of gain $\epsilon^\rho_\varphi$ usually increases (decreases) with congestion, resulting in ${\partial \epsilon^\lambda}/{\partial \varphi}<0$ (${\partial \epsilon^\lambda}/{\partial \varphi}>0$).
Although this relationship is established under the capacity sharing scenario, other service models, e.g., M/M/1 queue, can be studied in the same way and have the similar conclusion about the impact of traffic type on the elasticity of throughput\footnote{Interested readers are referred to Appendix \ref{appendix:a} for details of the case of M/M/1 queuing delay.}.
Based on the above discussions, the results of Corollary \ref{corollary:profit capacity} and \ref{corollary:welfare capacity} can provide important implications for APs and regulators to choose pricing strategies and regulatory policies under expanded system capacities.

\textbf{Summary of Implications:} From Corollary \ref{corollary:profit capacity} and \ref{corollary:welfare capacity}, whether an AP would decrease or increase its prices and whether regulators should relax or enhance price regulation largely depends on the changing direction of the elasticity of system throughput with changing congestion, which is influenced by the type of data traffic. In particular, in the early years of the Internet, data traffic was mainly for text content under which the elasticity of throughput often decreases with congestion. Under this case, when APs expand their capacities, they would lower the two-sided prices and regulators should relax the price regulation on the side of high market power. 
However, in recent years, data traffic is mostly for online video streaming under which the elasticity of throughput usually increases with congestion. Under this case, APs would increase the prices on both sides with expanded capacity, while regulators might want to tighten the price regulation on the side of high market power. 

\subsection{Impact of Users' Sensitivity}

In this subsection, we study the sensitivity of optimal pricing under varying users' sensitivity to congestion.
To model how sensitive end-users are to network congestion, we extend the gain function to be \(\rho(\phi,s)\), where the parameter \(s\) measures the congestion sensitivity of users.
Because when users become more sensitive to congestion, their throughput gain decreases more sharply with deteriorated congestion, we assume that \(\partial \rho(\phi,s_1)/\partial \phi>\partial \rho(\phi,s_2)/\partial \phi\) for all \(s_1<s_2\), which indicates that under any fixed level of congestion \(\phi\), if users become more sensitive to congestion, the marginal change in their throughput gain \(|\partial \rho/\partial \phi|\) increases.
Besides, we assume that the inverse \(\Lambda(\phi,\mu)\) of the congestion function satisfies \(\partial \Lambda(\phi,\mu_1)/\partial \phi \le \partial \Lambda(\phi,\mu_2)/\partial \phi\) for all \(\mu_1<\mu_2\), which intuitively states that if the AP's capacity is more abundant, the marginal change in the implied throughput \(\partial \Lambda/\partial \phi\) will not decrease. Both the congestion functions \(\Phi = \lambda/\mu\) and \(\Phi = 1/(\mu-\lambda)\) used before satisfy this assumption.
The following two corollaries show the impacts of the congestion sensitivity \(s\) of users on the profit- and welfare-optimal prices, respectively.

\begin{corollary}\label{corollary:profit sensitivity}
If the elasticity of system throughput increases with congestion, i.e., \(\partial \epsilon^\lambda/\partial \varphi>0\), the prices \(p^*\) and \(q^*\) both increase with the users' sensitivity to congestion \(s\), i.e.,
\[\frac{\partial p^*}{\partial s}>0\ \ \text{and} \ \ \frac{\partial q^*}{\partial s}>0.\]
Furthermore, it satisfies that
\begin{equation}\label{equation:proportion sensitivity}
\frac{\partial p^*}{\partial s}: \frac{\partial q^*}{\partial s} = \frac{d \tilde n^q}{d q}:\frac{d \tilde m^p}{d p}.
\end{equation}
\end{corollary}

Corollary \ref{corollary:profit sensitivity} states that if the elasticity of throughput $\epsilon^\lambda$ increases with congestion $\varphi$, the profit-optimal prices \(p^*\) and \(q^*\) both increase with the users' sensitivity \(s\).
As explained before, the monotonicity condition \(\partial \epsilon^\lambda/\partial \varphi>0\) often holds when network throughput is mostly constituted of inelastic traffic, e.g., online video streaming. This result implies that as video traffic continues to grow, users will become more sensitive to congestion and as a result, APs are expected to increase the prices on both sides so as to optimize their profits, which also leads to alleviated congestion and improved users' experiences.
Similar to Equation (\ref{equation:proportion capacity}), Equation (\ref{equation:proportion sensitivity}) shows that the marginal profit-optimal prices (with respect to sensitivity) are proportional to the marginal demand hazard rates (with respect to price) on the opposite sides.

\begin{corollary}\label{corollary:welfare sensitivity}
If the elasticity of system throughput increases with congestion, i.e., \(\partial \epsilon^\lambda/\partial \varphi>0\), the derivatives of the prices \(p^\circ\) and \(q^\circ\) with respect to the users' congestion sensitivity \(s\) satisfy that
\begin{align*}
\sgn \Big(\frac{\partial p^\circ}{\partial s}\Big) = -\sgn \Big(\frac{\partial q^\circ}{\partial s}\Big) = \sgn \left(\tilde m^p - \tilde n^q\right).
\end{align*}
\end{corollary}

Corollary \ref{corollary:welfare sensitivity} shows that under the monotonicity condition \(\partial \epsilon^\lambda/\partial \varphi>0\), if the demand hazard rate of the user side \(\tilde m^p\) is higher than that of the CP side \(\tilde n^q\), the derivate of \(p^\circ\) (\(q^\circ\)) with respect to the congestion sensitivity \(s\) is positive (negative), and vice-versa.
This result implies that when users become more sensitive to congestion, the welfare-optimal price of the side whose demand hazard rate is higher (lower) would increase (decrease).
By comparing the results of Corollary \ref{corollary:profit sensitivity} and \ref{corollary:welfare sensitivity}, we find that when the users' sensitivity to congestion increases, the price of the side of lower demand hazard rate needs to be reduced under the welfare-optimal pricing, but would be raised under the profit-optimal counterpart. This suggests that, to protect social welfare, when users become more sensitive to congestion, more stringent price regulation might need to be imposed on the side where the AP has a higher market power and lower demand hazard rate.

\textbf{Summary of Implications:} Corollary \ref{corollary:profit sensitivity} and \ref{corollary:welfare sensitivity} show how APs and regulators should adjust pricing strategies and regulatory policies under increasing congestion sensitivity of users, when data traffic is mostly for inelastic applications.
In particular, as video traffic keeps growing rapidly, end-users will become more sensitive to the network congestion, and consequently APs would increase the two-sided prices to alleviate the congestion and improve users' experiences so as to maximize profits. From a perspective of social welfare, regulators might want to impose more stringent price regulation on the side of high market power.

\section{Evaluation of Optimal Pricing}
\label{sec:evaluation}

In the previous sections, we studied the structures and sensitivities of the profit-optimal and welfare-optimal pricing through theoretical analysis. In this section, we further evaluate the pricing schemes by numerical simulations\footnote{Presently, APs usually implement the two-sided schemes in the form of paid peering or sponsored data plan on the CP side, whose pricing information has remained trade secrets. So there is no public data trace yet to evaluate our two-sided pricing model.}.

\subsection{Setup of Model Parameters}

We first choose the forms of congestion function \(\Phi(\lambda,\mu)\) and gain function \(\rho(\phi,s)\) to capture detailed characteristics of network services.
In particular, we adopt the congestion functions \(\Phi(\lambda,\mu) =\lambda/\mu\) and \(\Phi(\lambda,\mu) = 1/(\mu-\lambda)\). The former models the capacity sharing \cite{chau2010viability} nature of network services and was used in much prior work \cite{gibbens2000internet,jain2001analysis}; the latter models the M/M/1 queueing delay, which was also widely used in prior work \cite{ros2004mathematical,chau2010viability}.
We adopt the gain functions \(\rho(\phi,s) = 1/(\phi s+1)\) and \(\rho(\phi,s) = (s+1)^{-\phi}\) for \(s>0\), which were used in prior work \cite{richard2014pay,ma2013public}.
Their congestion elasticities are \(\epsilon^\rho_\phi = 1 - 1/(\phi s +1)\) and \(\epsilon^\rho_\phi = \phi \ln(s +1)\) and both increase with the congestion sensitivity of users \(s\). This indicates that the throughput gains will be more elastic to congestion if users become more sensitive to congestion.

\begin{figure*}[t]
 \centering
 \subfigure[\(\Phi= \frac{\lambda}{\mu}, \rho = \frac{1}{s\phi+1}\)]{
 \includegraphics[width=0.225\textwidth]{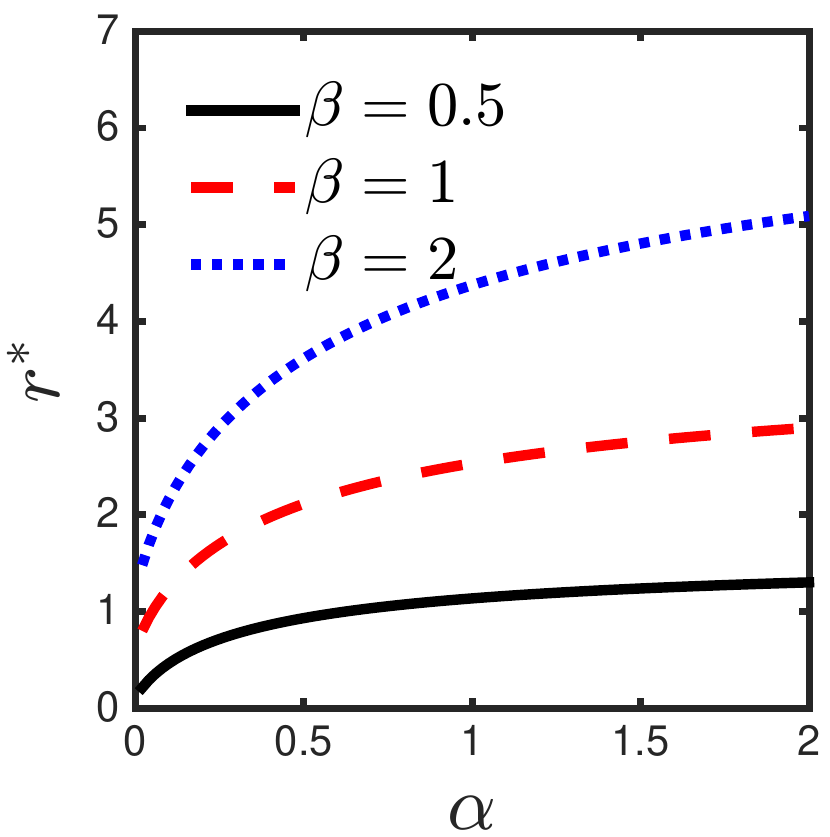}
 \label{figure:profit ratio sharing a}
 }
 \subfigure[\(\Phi= \frac{\lambda}{\mu},\rho = \frac{1}{(s+1)^{\phi}}\)]{
 \includegraphics[width=0.223\textwidth]{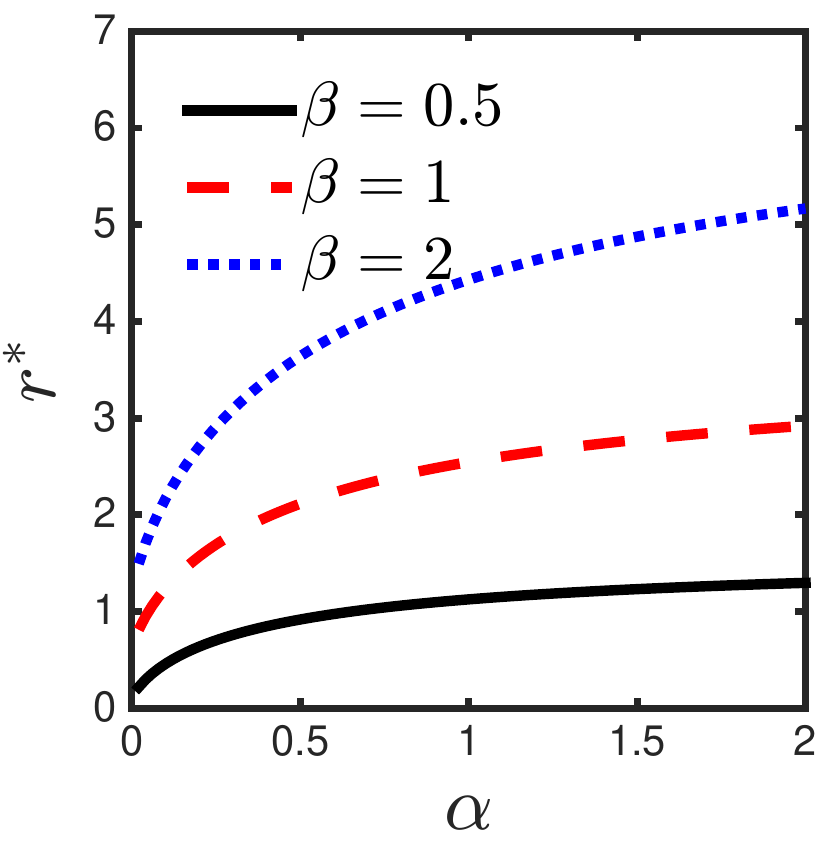}
 \label{figure:profit exponential a}
 }
  \subfigure[\(\Phi= \frac{1}{\mu-\lambda},\rho = \frac{1}{s\phi+1}\)]{
 \includegraphics[width=0.223\textwidth]{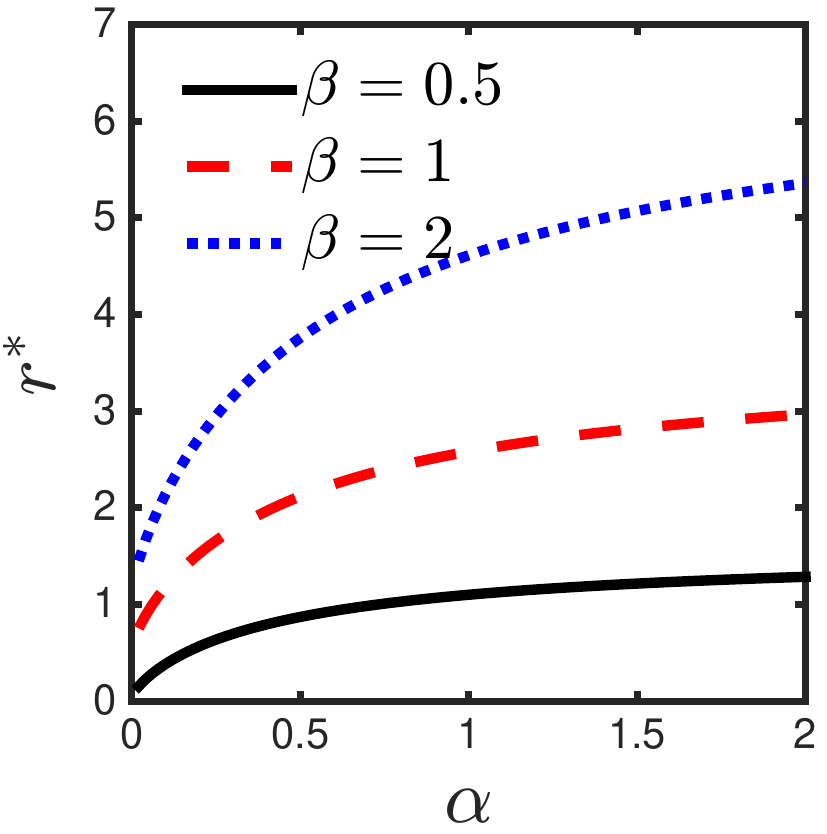}
 \label{figure:profit ratio mm1 a}
 }
 \vspace{-0.7em}
 \caption{\(r^*\) under varying \(\alpha\) and different \(\beta\).}
 \label{figure:profit ratio sharing}
\end{figure*}

\begin{figure*}[t]
 \centering
 \subfigure[\(\Phi= \frac{\lambda}{\mu}, \rho = \frac{1}{s\phi+1}\)]{
 \includegraphics[width=0.233\textwidth]{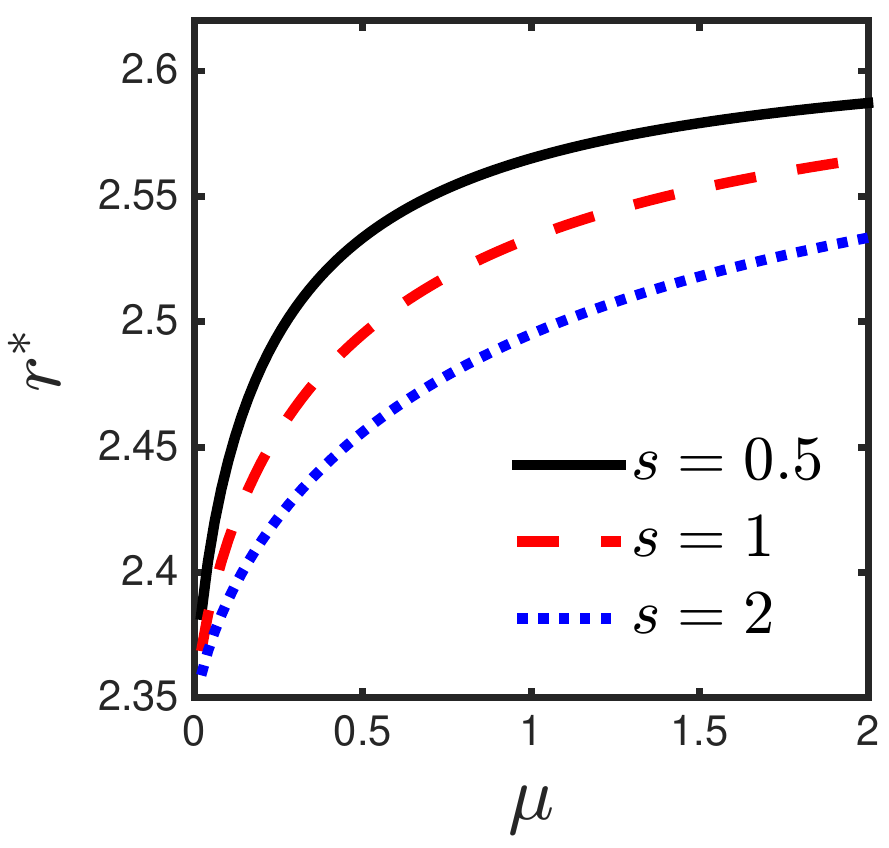}
 \label{figure:profit ratio sharing b}
 }
  \subfigure[\(\Phi= \frac{\lambda}{\mu},\rho = \frac{1}{(s+1)^{\phi}}\)]{
 \includegraphics[width=0.233\textwidth]{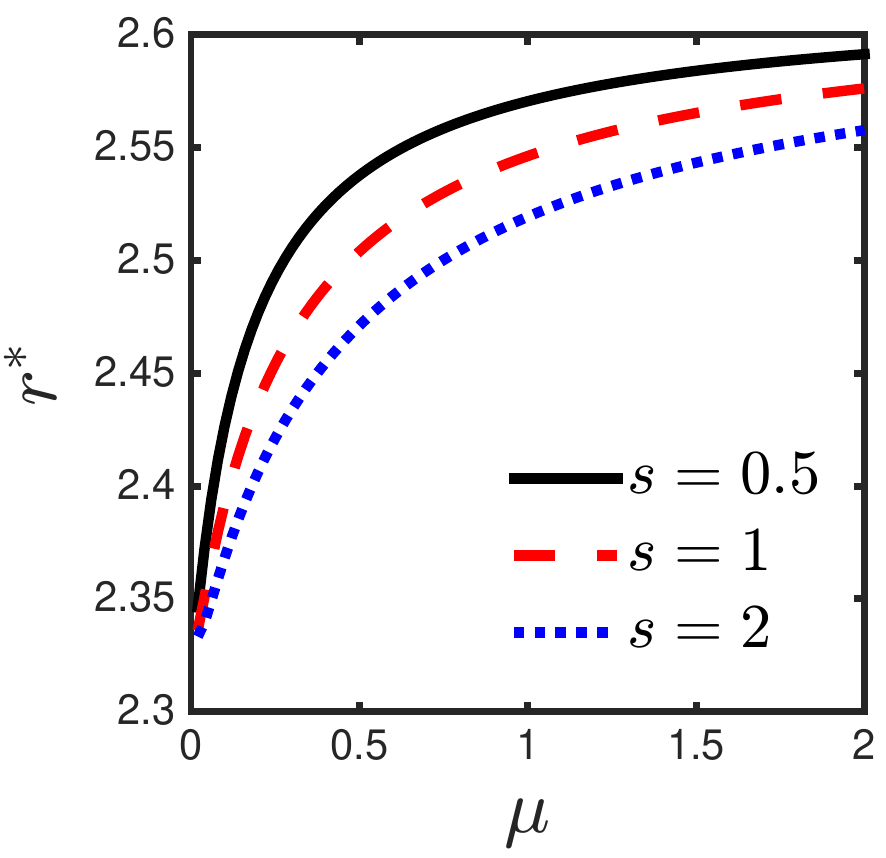}
 \label{figure:profit exponential b}
 }
  \subfigure[\(\Phi= \frac{1}{\mu-\lambda},\rho = \frac{1}{s\phi+1}\)]{
 \includegraphics[width=0.231\textwidth]{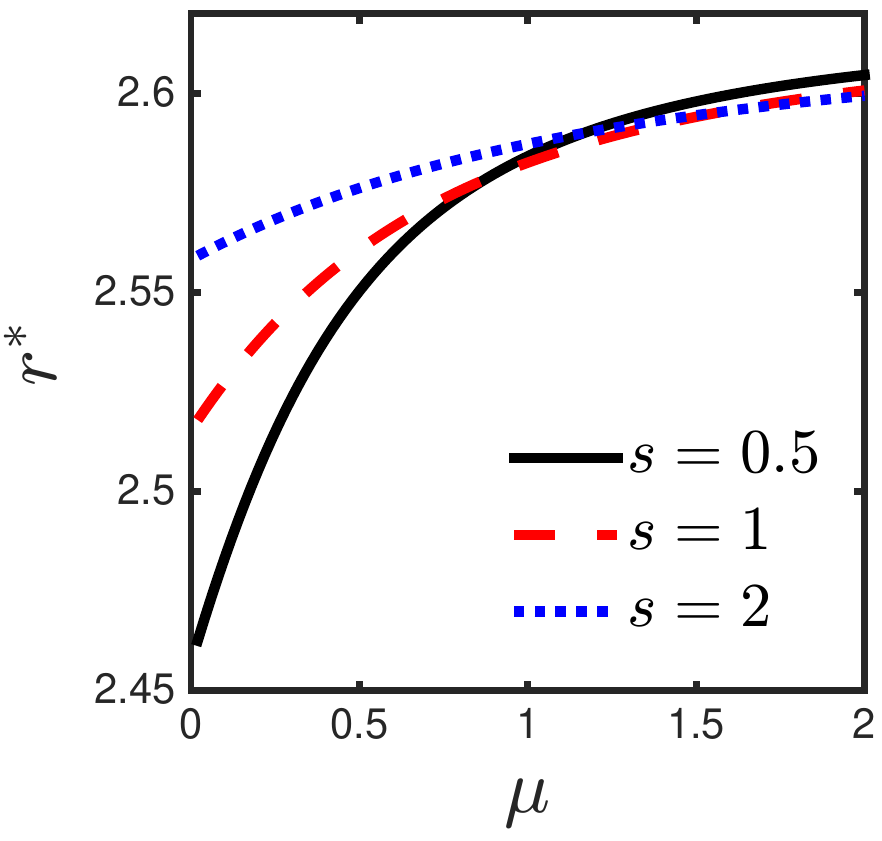}
 \label{figure:profit ratio mm1 b}
 }
 \vspace{-0.7em}
 \caption{\(r^*\) under varying \(\mu\) and different \(s\).}
 \label{figure:profit ratio mm1}
\end{figure*}

Although our analyses have focused on a single AP, market competition among multiple APs could be captured by the function \(m(p)\) of user population. In particular, we choose a family of population functions \(m\) parameterized by \(\alpha\): \(m(p,\alpha) \triangleq 1 - p^{\frac{1}{\alpha}}\) for \(0\le p \le 1,\alpha>0\), which satisfies that \(m(p,\alpha_1)>m(p,\alpha_2)\) for all \(\alpha_1<\alpha_2\). The parameter \(\alpha\) can be regarded as a metric of competition in the user market, i.e., given the same price \(p\), if the competition level \(\alpha\) increases, the user population \(m\) will fall.

The users' desirable throughput may change rapidly as the traffic demand of content services change.
For example, a new SuperHD video format launched by Netflix requires a \(50\%\) increase in traffic flows per video over 1080p content \cite{reed2014current}.
Similarly, we extend the function \(n(q)\) of average desirable throughput to capture this dynamics. In particular, we choose a family of throughput function \(n\) parameterized by \(\beta\): \(n(q,\beta) \triangleq 1-q^\beta\) for \(0\le q \le 1,\beta>0\), which satisfies $n(q,\beta_1) < n(q,\beta_2)$ for all $\beta_1 < \beta_2$.
The parameter $\beta$ can be regarded as a metric of traffic demand of content services, i.e., given the same price $q$, if the traffic demand $\beta$ increases, the desirable throughput will increase.

In the following simulations, we will compare various scenarios with a static baseline with \(\mu = s = \alpha=\beta=1\) and \(c=0.7\), under which the AP's capacity, the users' congestion sensitivity, the level of market competition and the traffic demand of content services are all normalized to one.

\subsection{Comparison with One-Sided Pricing}

Now we evaluate the AP's incentive to adopt the two-sided pricing instead of the traditional one-sided pricing that only charges on the user side. To this end, we denote the AP's profits by \(U^*_{one}\) and \(U^*_{two}\) under the profit-optimal one-sided and two-sided pricing, respectively. We define the growth rate of the profit by the two-sided pricing over the one-sided pricing by \(r^* \triangleq (U^*_{two} - U^*_{one})/U^*_{one}\).
A larger value of \(r^*\) corresponds to a higher profit growth for the AP; and therefore, leads to a stronger incentive for the AP to adopt the two-sided scheme.

\begin{figure}[t]
 \centering
 \subfigure[\(\Phi= \frac{1}{\mu-\lambda},\rho = \frac{1}{s\phi+1}\)]{
 \includegraphics[width=0.233\textwidth]{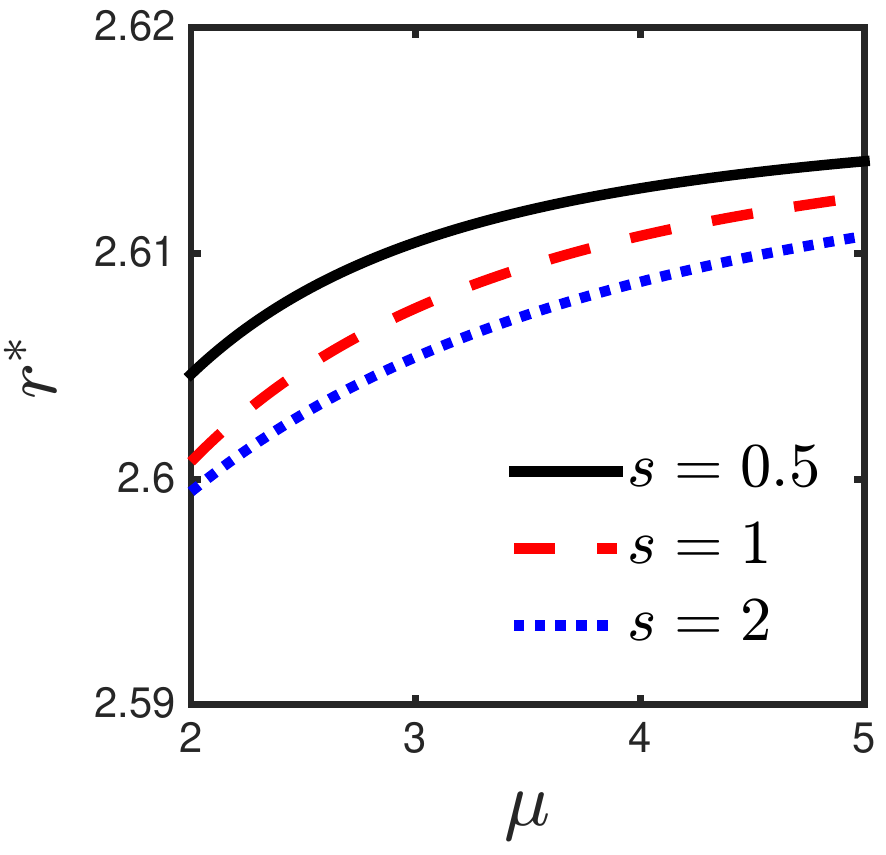}
 \label{figure:other a}
  }
 \subfigure[\(\Phi= \frac{1}{\mu-\lambda},\rho = \frac{1}{s\phi+1}\)]{
 \includegraphics[width=0.233\textwidth]{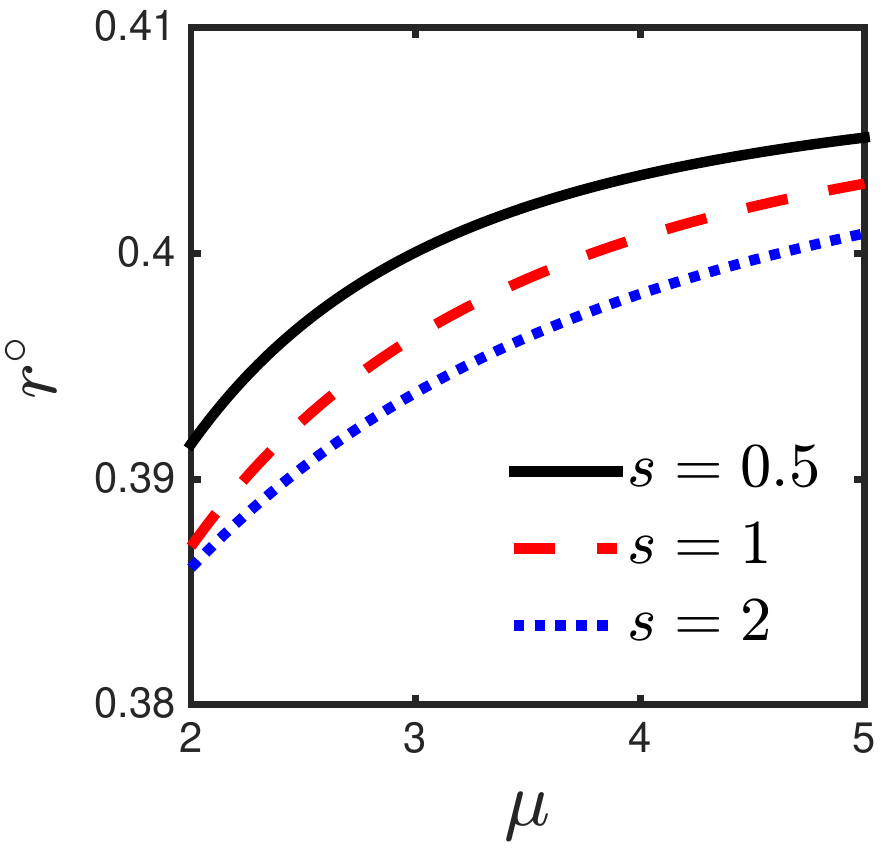}
 \label{figure:other b}
  }
 \caption{\(r^*\) and \(r^\circ\) under varying \(\mu>2\).}
 \label{figure:other}
\end{figure}

Figures \ref{figure:profit ratio sharing} and \ref{figure:profit ratio mm1} plot the profit growth rate \(r^*\) as a function of the competition level \(\alpha\) and the AP's capacity \(\mu\) under different traffic demand \(\beta\) and users' sensitivity \(s\), respectively.
The subfigures (a) are under the congestion function \(\Phi = \lambda/\mu\) and the gain function \(\rho = 1/(\phi s+1)\). From Subfigures (a) to (b), the gain function is changed to \(\rho = (s+1)^{-\phi}\). From Subfigures (a) to (c), the congestion function is changed to \(\Phi= 1/(\mu-\lambda)\).
In Figure \ref{figure:profit ratio sharing}, we observe that \(r^*\) increases with \(\alpha\) when the competition level on the user side becomes more intense, and larger values of the traffic demand \(\beta\) induce higher values of \(r^*\).
In Figure \ref{figure:profit ratio mm1}, we observe that \(r^*\) increases with \(\mu\) when the AP expands its capacity. Although \(r^*\) decreases with the sensitivity \(s\) in general, it increases with \(s\) under small capacity \(\mu\) when \(\Phi = 1/(\mu-\lambda)\). Because under the M/M/1 queueing delay \(1/(\mu-\lambda)\), the network congestion would be severe if the capacity is scarce. The increase of the users' sensitivity to congestion makes the network service more valuable for both users and CPs; and therefore, the AP can obtain a higher profit growth by adopting the two-sided pricing.
In fact, the non-monotonic trends in \(r^*\) happen only when the capacity \(\mu\) is scarce. To see this more clearly, we plot the case of \(\mu>2\) in Figure \ref{figure:other a}. From all of Figures \ref{figure:profit ratio sharing b}, \ref{figure:profit exponential b} and \ref{figure:other a}, we observe that a high value of \(s\) induces a lower value of \(r^*\).
In summary, our observations imply that the AP gets a higher profit growth and has stronger incentive to adopt the two-sided pricing if 1) the AP expands its capacity, 2) the competition level on the user side becomes fiercer, and 3) the content services require higher traffic throughput.

\begin{figure*}[t]
 \centering
 \subfigure[\(\Phi= \frac{\lambda}{\mu}, \rho = \frac{1}{s\phi+1}\)]{
 \includegraphics[width=0.23\textwidth]{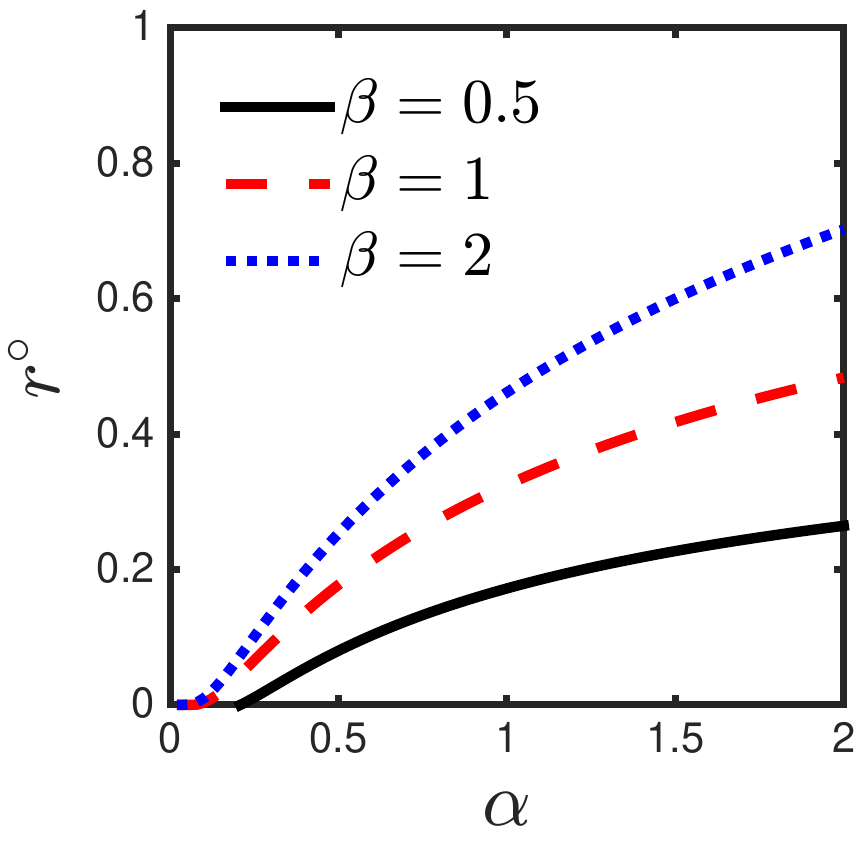}
  \label{figure:welfare ratio sharing a}
 }
  \subfigure[\(\Phi= \frac{\lambda}{\mu},\rho = \frac{1}{(s+1)^{\phi}}\)]{
 \includegraphics[width=0.23\textwidth]{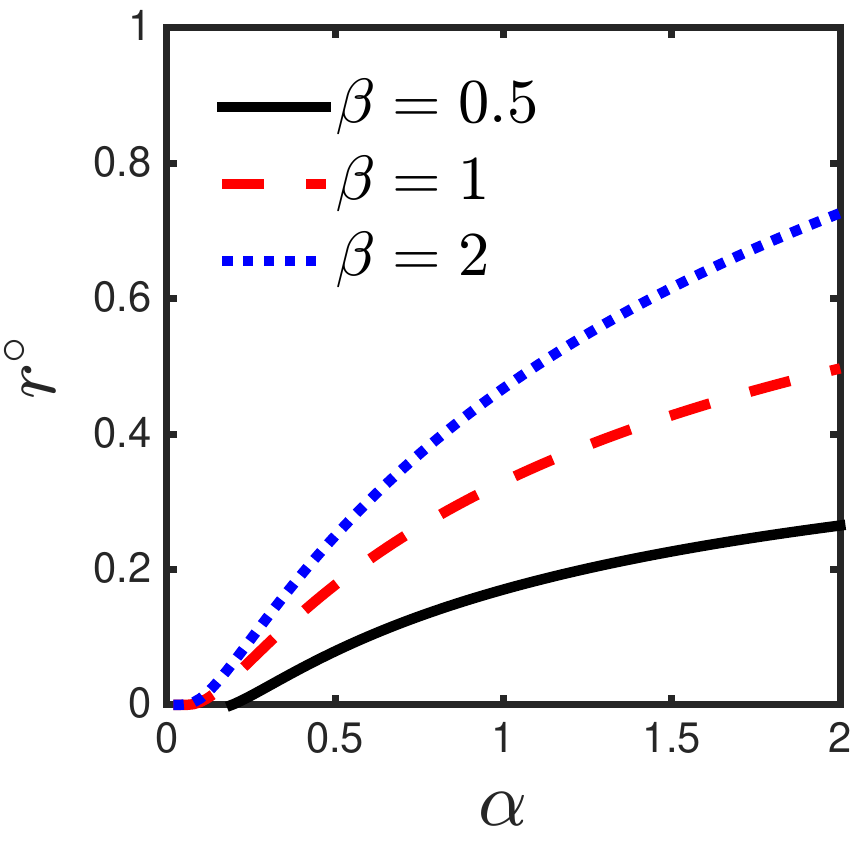}
 \label{figure:welfare exponential a}
 }
 \subfigure[\(\Phi= \frac{1}{\mu-\lambda},\rho = \frac{1}{s\phi+1}\)]{
 \includegraphics[width=0.23\textwidth]{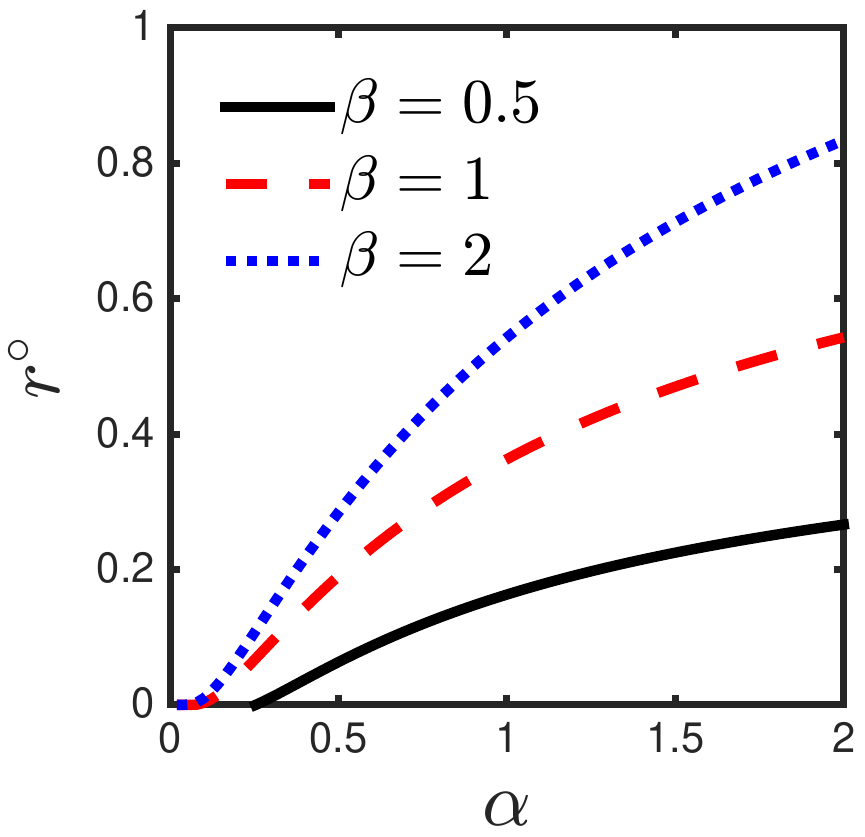}
 \label{figure:welfare ratio mm1 a}
  }
  \vspace{-0.7em}
 \caption{\(r^\circ\) under varying \(\alpha\) and different \(\beta\).}
 \label{figure:welfare ratio sharing}
\end{figure*}

\begin{figure*}[t]
 \centering
 \subfigure[\(\Phi= \frac{\lambda}{\mu}, \rho = \frac{1}{s\phi+1}\)]{
 \includegraphics[width=0.233\textwidth]{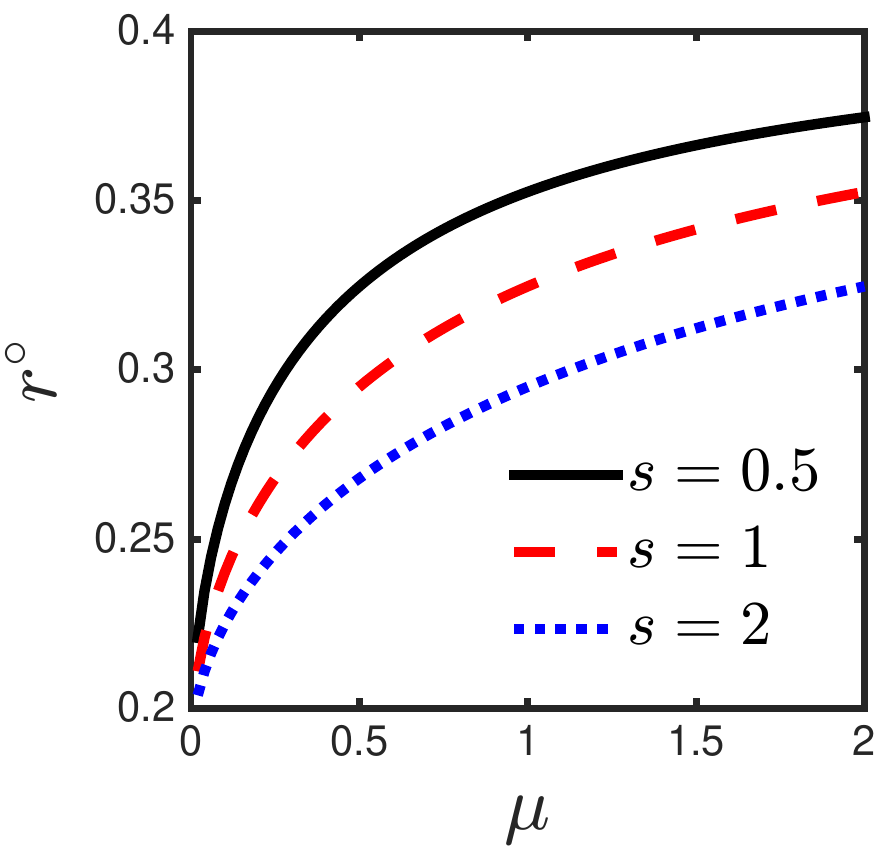}
  \label{figure:welfare ratio sharing b}
 }
 \subfigure[\(\Phi= \frac{\lambda}{\mu},\rho = \frac{1}{(s+1)^{\phi}}\)]{
 \includegraphics[width=0.228\textwidth]{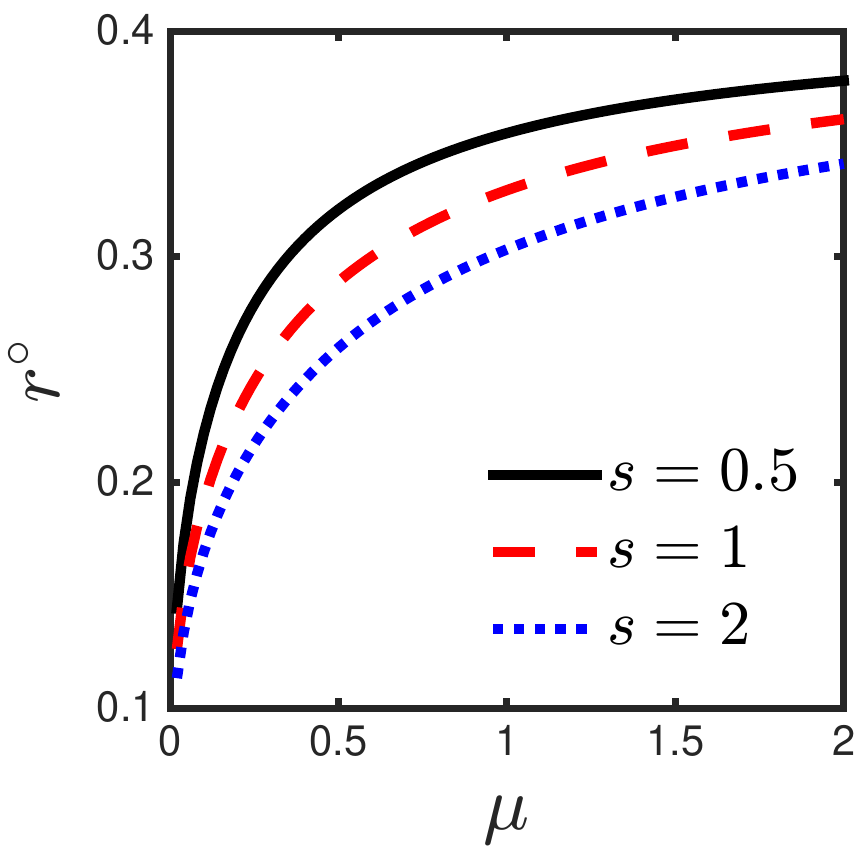}
 \label{figure:welfare exponential b}
 }
 \subfigure[\(\Phi= \frac{1}{\mu-\lambda},\rho = \frac{1}{s\phi+1}\)]{
 \includegraphics[width=0.234\textwidth]{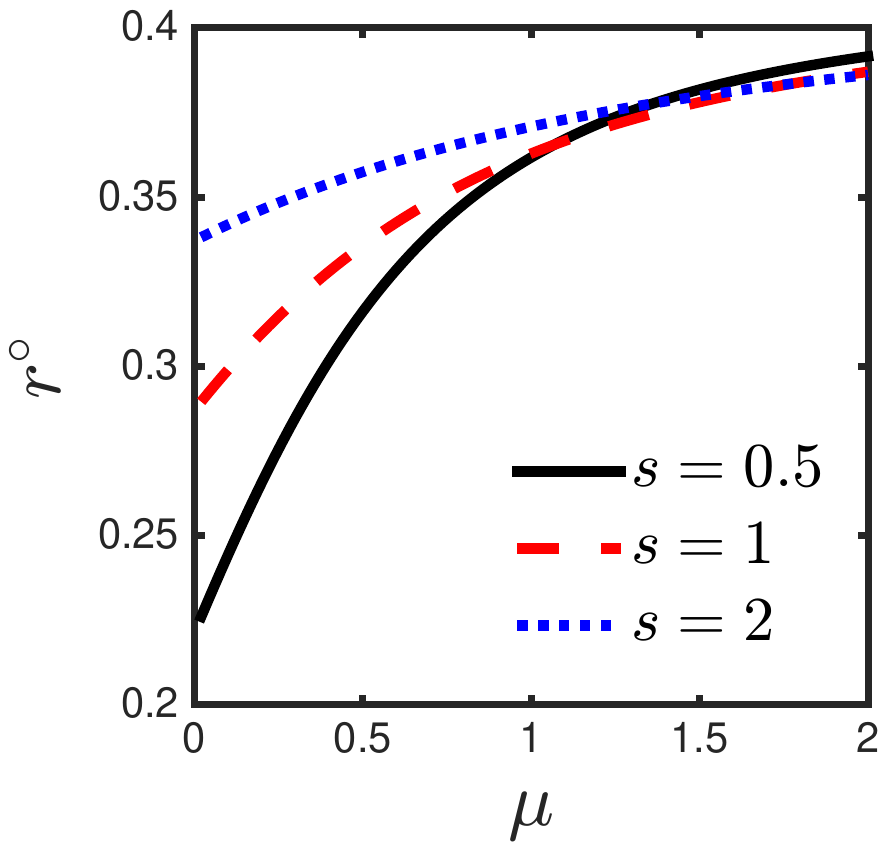}
 \label{figure:welfare ratio mm1 b}
  }
  \vspace{-0.7em}
 \caption{\(r^\circ\) under varying \(\mu\) and different \(s\).}
 \label{figure:welfare ratio mm1}
\end{figure*}

Next, we evaluate how regulators should deal with the shift from the one-sided pricing to the two-sided pricing. For this purpose, we denote the social welfare under the welfare-optimal one-sided and two-sided pricing
by \(W^\circ_{one}\) and \(W^\circ_{two}\), respectively. Similarly, we define the growth rate of social welfare by the two-sided pricing over the one-sided pricing by \(r^\circ \triangleq (W^\circ_{two}-W^\circ_{one})/W^\circ_{one}\). A larger value of \(r^\circ\) indicates that social welfare gets a higher growth when shifting from the one-sided to the two-sided pricing, and thus regulators might want to encourage this transformation.

Figures \ref{figure:welfare ratio sharing} and \ref{figure:welfare ratio mm1} plot the welfare growth rate \(r^\circ\) as a function of the competition level \(\alpha\) and the AP's capacity \(\mu\) under different traffic demand \(\beta\) and users' sensitivity \(s\), respectively.
As a complement to Figure \ref{figure:welfare ratio mm1 b}, Figure \ref{figure:other b} plots the case of large capacity under the M/M/1 queueing setting.
By comparing Figures \ref{figure:welfare ratio sharing}, \ref{figure:welfare ratio mm1} and \ref{figure:other b} with Figures \ref{figure:profit ratio sharing}, \ref{figure:profit ratio mm1} and \ref{figure:other a}, respectively, we observe that the welfare growth rate \(r^\circ\) has the same changing trend as the profit growth rate \(r^*\) when the parameter \(\mu,s,\alpha\) or \(\beta\) varies. This observation provides justifications for regulators to encourage the AP to shift from the one-sided to the two-sided pricing, especially when the AP has strong incentive to do so.

\subsection{Impacts of Competition and Demand}

\begin{figure}[]
 \centering
 \space\space\space\space
 \includegraphics[width=0.205\textwidth]{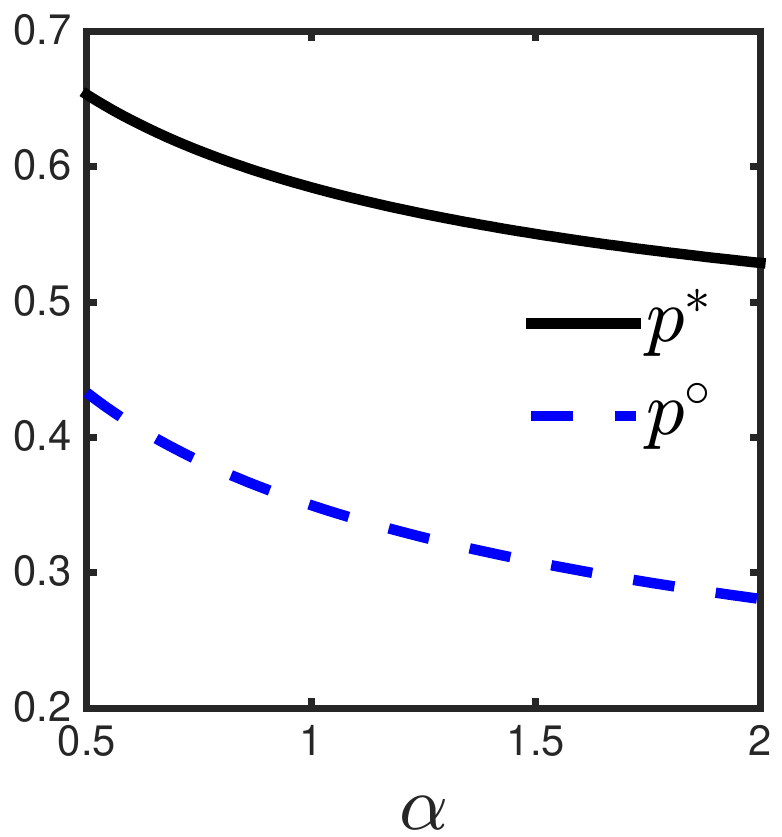}
 \space\space\space\space
 \includegraphics[width=0.205\textwidth]{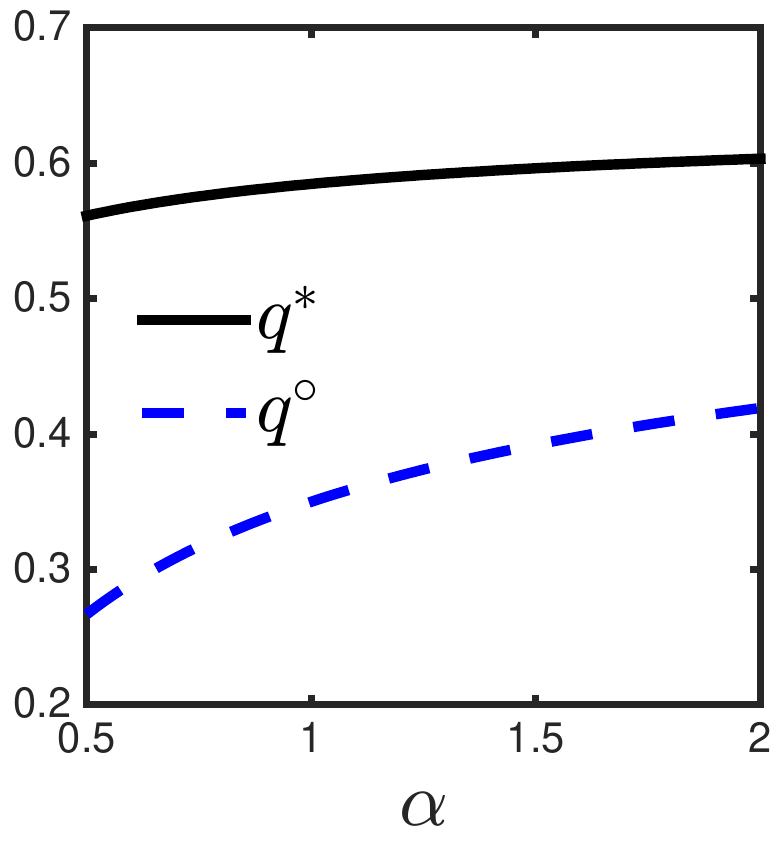}
 \space\space\space\space
 \includegraphics[width=0.205\textwidth]{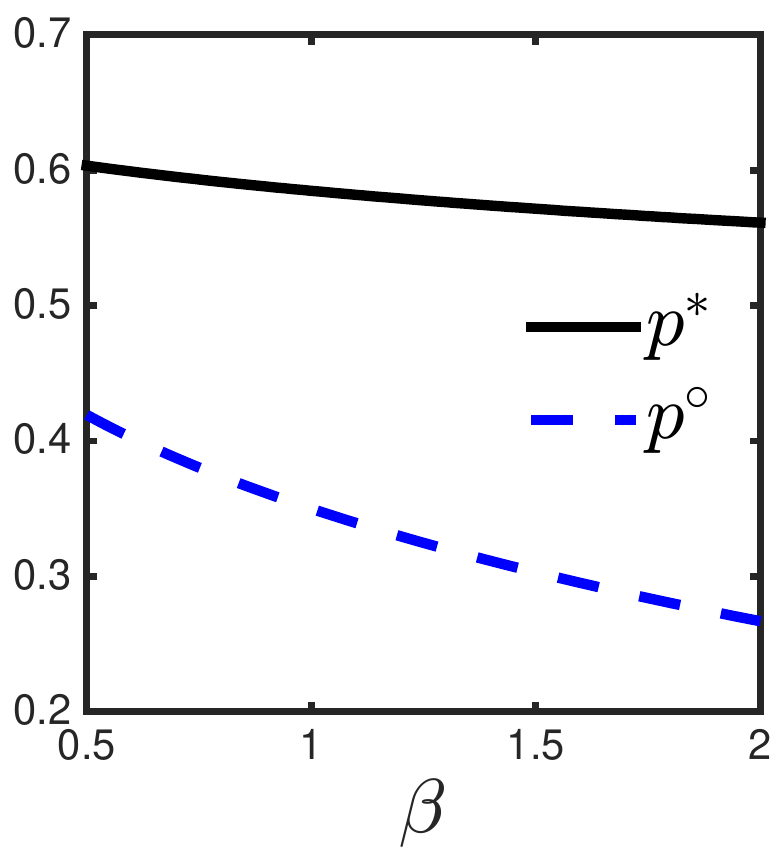}
 \space\space\space\space
 \includegraphics[width=0.205\textwidth]{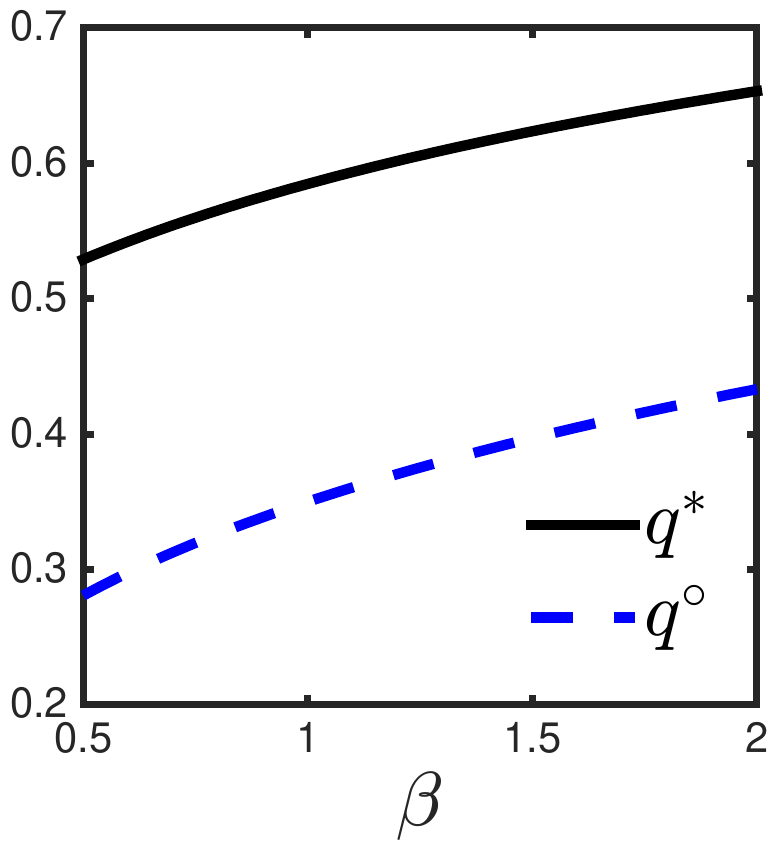}
 \caption{The profit-optimal and welfare-optimal pricing schemes under varying \(\alpha\) and \(\beta\).}
 \label{figure:varying_capacitysharing}
\end{figure}

Since we have studied the changes of the optimal pricing under varying capacity \(\mu\) and sensitivity \(s\) and provided analytical results in the previous section, we next focus on understanding the impacts of the competition level \(\alpha\) and traffic demand \(\beta\) on the optimal prices.
Figure \ref{figure:varying_capacitysharing} plots the profit-optimal prices \((p^*,q^*)\) and the welfare-optimal prices \((p^\circ,q^\circ)\) as functions of \(\alpha\) and \(\beta\) under the congestion function \(\Phi = \lambda/\mu\) and the gain function \(\rho = 1/(\phi s+1)\). From Figure \ref{figure:varying_capacitysharing}, we observe that 1) when \(\alpha\) or \(\beta\) increases, the user-side prices \(p^*,p^\circ\) decrease and the CP-side prices \(q^*,q^\circ\) increase,
and 2) the welfare-optimal prices \(p^\circ\) and \(q^\circ\) are always lower than the profit-optimal prices \(p^*\) and \(q^*\), respectively.
The first observation implies that as the APs' competition in the user market becomes more intense or the content services have larger traffic demand, the user-side price will decrease and the CP-side price will increase, regardless whether the objective is the AP's profit or social welfare.
The second observation indicates that, to optimize social welfare, regulators might want to regulate the prices on both the user and CP sides, which coincides with the implication of Proposition \ref{proposition:welfare}. 
Note that we omit the cases when the congestion function is \(\Phi = 1/(\mu-\lambda)\) or/and the gain function is \(\rho = (s+1)^{-\phi}\). Because in the cases, the changing trends of the optimal prices under varying \(\alpha\) and \(\beta\) are the same as those shown in Figure \ref{figure:varying_capacitysharing}.

\textbf{Summary of Implications}: Our observations in this section imply that as the capacities of APs and the demand for video traffic grow in the current Internet, APs will have increasing incentives to transform from the traditional one-sided pricing on the user side to the two-sided pricing. Under these cases, regulators might want to encourage this transformation, since it will bring higher growth rates for both social welfare and APs' profits. 

\section{Conclusions}
In this paper, we study the optimal two-sided pricing for congested networks.
We present a novel model to capture end-users' population and throughput demand under pricing and congestion parameters and derive the system congestion under an equilibrium.
Based on this model, we characterize the structures of the profit-optimal and welfare-optimal pricing schemes.
Our results reveal that the profit-optimal pricing always equalizes the demand hazard rates at both sides, while the welfare-optimal counterpart differentiates them based on the elasticity of system throughput.
We also explore the sensitivities of the optimal pricing under varying system parameters.
We find that with the growth of APs' capacities and end-users' traffic demand in the current Internet, APs will have increasing incentives to shift from the traditional one-sided pricing to the two-sided pricing, because it brings higher growth rates for their profits. Furthermore, as online video streaming becomes the main source of network traffic and keeps growing rapidly, end-users become more sensitive to the network congestion and APs are expanding their capacities. Under such scenarios, APs may increase two-sided prices to alleviate the congestion and improve users' experiences so as to maximize profits. From the perspective of social welfare, regulators might want to tighten the price regulation on the side where APs have higher market power and lower demand hazard rate.

\appendix

\section{The Case of M/M/1 queue}
\label{appendix:a}
In Section \ref{section:5.1}, we showed that for the capacity sharing scenario, we have the result that if the data traffic is for text (online video), the elasticity of throughput \(\epsilon^\lambda\) decreases (increases) with congestion under a reasonable condition. The condition is that the congestion elasticity of throughput gain of text (online video) traffic increases (decreases) with congestion.
In this section, we show that for the congestion model of M/M/1 queue, we can derive the similar result under another reasonable condition.

For the scenario of M/M/1 queue, the congestion function is \(\Phi(\lambda,\mu) = 1/(\mu-\lambda)\) and its inverse function with respect to the throughput \(\lambda\) is \(\Lambda(\phi,\mu) = \mu - 1/\phi\). By Equation (\ref{equation:system elasticity}), the elasticity of system throughput \(\epsilon^\lambda\) can be expressed as
\begin{equation}
\epsilon^\lambda(\varphi) = \frac{1}{1+mnG(\varphi)}
\label{equation:MM1}
\end{equation}
where \(G(\varphi) \triangleq -\varphi^2d\rho(\varphi)/d\varphi\).
By Equation (\ref{equation:MM1}), the elasticity of throughput \(\epsilon^\lambda\) decreases with \(G\).
When the data traffic is for text, the throughput gain $\rho(\varphi)$ usually decreases concavely in congestion $\varphi$, i.e., \(d^2 \rho(\varphi)/d \varphi^2<0\), as text traffic is insensitive to mild congestion. Under such a case, we have that \(G(\varphi)\) increases with congestion \(\varphi\), i.e., \(dG(\varphi)/d\varphi>0\). Conversely, when the data traffic is for online video, we consider that \(G(\varphi)\) often decreases with congestion \(\varphi\).
Therefore, under the condition that the function \(G(\varphi)\) for text (online video) traffic increases (decreases) with congestion, we have the conclusion that
if the data traffic is for text (online video), the elasticity of throughput \(\epsilon^\lambda\) decreases (increases) with congestion by Equation (\ref{equation:MM1}).

\section{Proofs of Theoretical Results}
In this section, we provide the proofs of the theoretical results mentioned in our paper.

\hspace{-0.12in}\textbf{Proof of Theorem \ref{theorem:unique-congestion}:} By Assumption \ref{ass:congestion_function}, \(\Phi(\lambda,\mu)\) is increasing in \(\lambda\) and thus its inverse function \(\Lambda(\phi,\mu)\) is increasing in \(\phi\). By Assumption \ref{ass:gain}, \(\lambda(\phi)\) is decreasing in \(\phi\) and therefore \(g(\phi) = \Lambda(\phi,\mu) - \lambda(\phi)\) is increasing in \(\phi\).
Because \(g(\phi)\) is a continuously increasing function of \(\phi\) and it satisfies that
\begin{align*}
& g\big(\Phi(0,\mu)\big) = \Lambda\big(\Phi(0,\mu),\mu\big) - \lambda\big(\Phi(0,\mu)\big) = - \lambda\big(\Phi(0,\mu)\big)<0 \quad \text{and}\vspace{0.06in}\\
&\displaystyle\lim_{\phi\rightarrow +\infty}  g(\phi) =   \lim_{\phi\rightarrow +\infty} \left[\Lambda(\phi,\mu)-\lambda(\phi)\right]= \lim_{\phi\rightarrow +\infty} \Lambda(\phi,\mu)>0,
\end{align*}
the equation \(g(\phi) = 0\) must have a unique solution on the interval \(\big[\Phi(0,\mu), +\infty\big)\) by the intermediate value theorem. Because the minimum of \(\phi\) is \(\Phi(0,\mu)\) under which the aggregate throughput \(\lambda\) is zero, there always exists a unique congestion \(\phi\) which solves \(g(\phi) = 0\), i.e., satisfies \(\phi = \Phi(\lambda(\phi),\mu)\) and is an equilibrium congestion by Definition \ref{def:congestion}.
\QEDA\\

\hspace{-0.12in}\textbf{Proof of Proposition \ref{proposition:elasticity}:} Under the equilibrium of the system \((m,n,\mu)\), \(\Lambda\) and \(\lambda\) are functions of \(m,n,\mu\) and we denote them by \(\Lambda(m,n,\mu) \triangleq \Lambda\left(\varphi(m,n,\mu),\mu\right)\) and \(\lambda(m,n,\mu) \triangleq \lambda\left(m,n,\varphi(m,n,\mu)\right) = mn\rho\left(\varphi(m,n,\mu)\right)\). Correspondingly, we denote the throughput gap by \(g(m,n,\mu) \triangleq \Lambda(m,n,\mu) - \lambda(m,n,\mu)\). By Theorem \ref{theorem:unique-congestion}, under the equilibrium, the throughput gap \(g(m,n,\mu)\) is zero for any \(m,n,\) and \(\mu\). Thus, for the user population \(m\), we have the identity that 
\[\frac{\partial g(m,n,\mu)}{\partial m} = \frac{\partial \Lambda(\varphi,\mu)}{\partial \varphi} \frac{\partial \varphi(m,n,\mu)}{\partial m} - \frac{\partial \lambda(m,n,\mu)}{\partial m} = 0 \ \ \text{where} \ \ \frac{\partial \lambda(m,n,\mu)}{\partial m} = \frac{\lambda(m,n,\varphi)}{\partial m} + \frac{\partial \lambda(m,n,\varphi)}{\partial \varphi}\frac{\partial \varphi(m,n,\mu)}{\partial m}.\]
Based on this identity, we can derive that
\[\frac{\partial \varphi(m,n,\mu)}{\partial m} = \left(\frac{\partial \Lambda(\varphi,\mu)}{\partial \varphi} - \frac{\partial \lambda(m,n,\varphi)}{\partial \varphi}\right)^{-1} \frac{\lambda(m,n,\varphi)}{\partial m} = \left(\frac{\partial g(m,n,\mu,\varphi)}{\partial \varphi}\right)^{-1}\frac{\lambda}{m} > 0\]
where we denote \(g(m,n,\mu,\varphi) \triangleq \Lambda(\varphi,\mu)-\lambda(m,n,\varphi)\). Furthermore, we have that
\[\frac{\partial \lambda(m,n,\mu)}{\partial m} =\frac{\partial \Lambda(\varphi,\mu)}{\partial \varphi} \frac{\partial \varphi(m,n,\mu)}{\partial m} > 0.\]
Similarly, for the average throughput \(n\), we can derive that
\[\frac{\partial \varphi(m,n,\mu)}{\partial n}= \left(\frac{\partial g(m,n,\mu,\varphi)}{\partial \varphi}\right)^{-1}\frac{\lambda}{n} > 0 \quad \text{and} \quad \frac{\partial \lambda(m,n,\mu)}{\partial n} =\frac{\partial \Lambda(\varphi,\mu)}{\partial \varphi} \frac{\partial \varphi(m,n,\mu)}{\partial n} > 0.\]
For the AP's capacity \(\mu\), we have the identity that
\[\frac{\partial g(m,n,\mu)}{\partial \mu} = \left(\frac{\partial \Lambda(\varphi,\mu)}{\partial \varphi} \frac{\partial \varphi(m,n,\mu)}{\partial \mu}+\frac{\partial \Lambda(\varphi,\mu)}{\partial \mu}\right) - \frac{\partial \lambda(m,n,\varphi)}{\partial \varphi}\frac{\partial \varphi(m,n,\mu)}{\partial \mu} = 0\]
from which we can derive that
\[\frac{\partial \varphi(m,n,\mu)}{\partial \mu}  = -\frac{\partial \Lambda(\varphi,\mu)}{\partial \mu}\left(\frac{\partial \Lambda(\varphi,\mu)}{\partial \varphi} - \frac{\partial \lambda(m,n,\varphi)}{\partial \varphi}\right)^{-1} = -\frac{\partial \Lambda(\varphi,\mu)}{\partial \mu}\left(\frac{\partial g(m,n,\mu,\varphi)}{\partial \varphi}\right)^{-1}<0.\]
Furthermore, it satisfies that
\[\frac{\partial \lambda(m,n,\mu)}{\partial \mu} = \frac{\partial \lambda(m,n,\varphi)}{\partial \varphi}\frac{\partial \varphi(m,n,\mu)}{\partial \mu} = mn\frac{d \rho(\varphi)}{d \varphi}\frac{\partial \varphi(m,n,\mu)}{\partial \mu}> 0.\tag*{\QEDA}\]

\hspace{-0.12in}\textbf{Proof of Theorem \ref{theorem:elasticity}:} Based on Definition \ref{def:elasticity} and Proposition \ref{proposition:elasticity}, we have the identities that
\begin{align*}
&\epsilon^\lambda_m = \frac{\partial \lambda(m,n,\mu)}{\partial m} \left(\frac{\lambda}{m}\right)^{-1} = \frac{\partial \Lambda(\varphi,\mu)}{\partial \varphi}\left(\frac{\partial g(m,n,\mu,\varphi)}{\partial \varphi}\right)^{-1}\quad \text{and} \\
&\epsilon^\lambda_n = \frac{\partial \lambda(m,n,\mu)}{\partial n} \left(\frac{\lambda}{n}\right)^{-1} = \frac{\partial \Lambda(\varphi,\mu)}{\partial \varphi}\left(\frac{\partial g(m,n,\mu,\varphi)}{\partial \varphi}\right)^{-1}
\end{align*}
where it satisfies that
\[\frac{\partial \Lambda(\varphi,\mu)}{\partial \varphi}\left(\frac{\partial g(m,n,\mu,\varphi)}{\partial \varphi}\right)^{-1}    =  \frac{{\partial \Lambda(\varphi,\mu)}/{\partial \varphi}}{{\partial \Lambda(\varphi,\mu)}/{\partial \varphi}  -  {\partial \lambda(m,n,\varphi)}/{\partial \varphi}}  = \left( 1 +  \frac{|\partial \lambda(m,n,\varphi)/\partial \varphi|}{{\partial \Lambda(\varphi,\mu)}/{\partial \varphi}} \right)^{  -1}   \in (0,1].\tag*{\QEDA}\]

\hspace{-0.12in}\textbf{Proof of Proposition \ref{proposition:pricing-effect}:} From Proposition \ref{proposition:elasticity}, we can derive that
\[\frac{\partial \varphi(p,q,\mu)}{\partial p} = \frac{\partial \varphi(m,n,\mu)}{\partial m}\frac{d m(p)}{d p} = \left(\frac{\partial g(m,n,\mu,\varphi)}{\partial \varphi}\right)^{-1} \frac{\lambda}{m} \frac{d m(p)}{d p} = -\left(\frac{\partial g(m,n,\mu,\varphi)}{\partial \varphi}\right)^{-1} \lambda \tilde m^p < 0,\]
and therefore we have that
\[\frac{\partial \lambda(p,q,\mu)}{\partial p} = \frac{\partial \Lambda(p,q,\mu)}{\partial p}  = \frac{\partial \Lambda(\varphi,\mu)}{\partial \varphi} \frac{\partial \varphi(p,q,\mu)}{\partial p} < 0\]
where we denote \(\Lambda(p,q,\mu) \triangleq \Lambda\left(\varphi(p,q,\mu),\mu\right)\).
Similarly, we can derive that
\[\frac{\partial \varphi(p,q,\mu)}{\partial q} = -\left(\frac{\partial g(m,n,\mu,\varphi)}{\partial \varphi}\right)^{-1} \lambda \tilde n^q < 0 \quad \text{and} \quad \frac{\partial \lambda(p,q,\mu)}{\partial q} = \frac{\partial \Lambda(\varphi,\mu)}{\partial \varphi} \frac{\partial \varphi(p,q,\mu)}{\partial q}<0.\]
Besides, by Theorem \ref{theorem:elasticity}, it satisfies that
\[\epsilon_p^\lambda:\epsilon_q^\lambda = \left(\epsilon_m^\lambda\epsilon_p^m\right) : \left(\epsilon_n^\lambda\epsilon_q^n\right)= \epsilon_p^m:\epsilon_q^n.\tag*{\QEDA}\]

\hspace{-0.12in}\textbf{Proof of Proposition \ref{proposition:profit}:} From Proposition \ref{proposition:elasticity} and Equation (\ref{equation:system elasticity}), the impact of the capacity \(\mu\) on the profit is
\begin{align*}
\frac{\partial U(p,q,\mu)}{\partial \mu} &= (p+q-c)\frac{\partial \lambda(p,q,\mu)}{\partial \mu} = (p+q-c)\frac{\partial \lambda(m(p),n(q),\mu)}{\partial \mu}\\
&= -(p+q-c)\frac{\partial \Lambda(\varphi,\mu)}{\partial \mu}mn\frac{d \rho(\varphi)}{\partial \varphi}\left(\frac{\partial g(m,n,\mu,\varphi)}{\partial \varphi}\right)^{-1}=(p + q -c)\frac{\partial \Lambda(\varphi,\mu)}{\partial \mu}(1-\epsilon^\lambda)>0.
\end{align*}
From Proposition \ref{proposition:pricing-effect} and Equation (\ref{equation:system elasticity}), the impacts of the two-sided prices on the profit are
\begin{align*}
\frac{\partial U(p,q,\mu)}{\partial p} &\!=\! \lambda \!+\! (p+q-c)\frac{\partial \lambda(p,q,\mu)}{\partial p} \!=\! \lambda \!-\! (p+q-c)\frac{\partial \Lambda(\varphi,\mu)}{\partial \varphi}\left(\frac{\partial g(m,n,\mu,\varphi)}{\partial \varphi}\right)^{-1} \!\!\lambda \tilde{m}^p \!=\! \lambda \!-\!(p+q-c)\epsilon^\lambda \lambda \tilde{m}^p, \\
\frac{\partial U(p,q,\mu)}{\partial q} &\!=\! \lambda \!+\! (p+q-c)\frac{\partial \lambda(p,q,\mu)}{\partial q} \!=\! \lambda \!-\! (p+q-c)\frac{\partial \Lambda(\varphi,\mu)}{\partial \varphi}\left(\frac{\partial g(m,n,\mu,\varphi)}{\partial \varphi}\right)^{-1} \!\!\lambda \tilde{n}^q \!=\! \lambda \!-\! (p+q-c)\epsilon^\lambda \lambda \tilde{n}^q.\tag*{\QEDA}
\end{align*}

\hspace{-0.12in}\textbf{Proof of Theorem \ref{theorem:KKT-Lerner}:} By the Karush-Kuhn-Tucker (KKT) necessary conditions and Proposition \ref{proposition:profit}, if the two-sided prices \((p,q)\) maximize the profit \(U\), we have the relations that
\begin{align}\label{equation:KKT proof}
\begin{cases}
\displaystyle\frac{\partial U(p,q,\mu)}{\partial p} = \lambda - (p+q-c)\epsilon^\lambda\lambda \tilde m^p = 0\vspace{0.03in}\\
\displaystyle\frac{\partial U(p,q,\mu)}{\partial q} = \lambda - (p+q-c)\epsilon^\lambda\lambda \tilde m^q = 0
\end{cases}
\end{align}
from which we can derive Equation (\ref{equation:KKT-necesary}), i.e.,
\[\tilde m^p = \tilde n^q = \frac{1}{(p+q-c)\epsilon^\lambda}.\]
Furthermore, by Definition \ref{def:elasticity} and \ref{def:hazard-rate}, it satisfies that
\begin{align*}
\begin{cases}
\displaystyle\epsilon^\lambda\epsilon^m_p = \epsilon^\lambda p\tilde m^p = \frac{p}{p+q-c}\vspace{0.03in}\\
\displaystyle\epsilon^\lambda\epsilon^n_q  = \epsilon^\lambda q\tilde n^q = \frac{q}{p+q-c}
\end{cases}
\end{align*}
from which the total price \(p+q\) satisfies that
\[\displaystyle\frac{p+q}{p+q-c}  = \epsilon^\lambda\epsilon^m_p + \epsilon^\lambda\epsilon^n_q= \epsilon^\lambda_p + \epsilon^\lambda_q\]
and thus Equation (\ref{equation:Lerner-formula}) holds.
\QEDA\\

\hspace{-0.12in}\textbf{Proof of Proposition \ref{proposition:welfare}:} From Proposition \ref{proposition:elasticity}, the impact of the capacity \(\mu\) on social welfare \(W\) is
\begin{align*}
\frac{\partial W(p,q,\mu)}{\partial \mu} &= (s_m+s_n)\frac{\partial \lambda(m(p),n(q),\mu)}{\partial \mu} = \frac{W}{\lambda}\frac{\partial \lambda(m(p),n(q),\mu)}{\partial \mu} \\
&= -\frac{W}{mn\rho(\varphi)}mn\frac{d\rho(\varphi)}{d\varphi}\frac{\partial \Lambda(\varphi,\mu)}{\partial \mu}\left(\frac{\partial g(m,n,\mu,\varphi)}{\partial \varphi}\right)^{-1} = W\tilde \rho^\varphi\frac{\partial \Lambda(\varphi,\mu)}{\partial \mu}\left(\frac{\partial g(m,n,\mu,\varphi)}{\partial \varphi}\right)^{-1} >0.
\end{align*}
From Proposition \ref{proposition:pricing-effect} and Equation (\ref{equation:system elasticity}), the impact of the user-side price \(p\) on social welfare \(W\) is
\begin{align*}
\displaystyle\frac{\partial W(p,q,\mu)}{\partial p}&= \frac{d s_m(p)}{d p}\lambda + \big(s_m + s_n\big)\frac{\partial \lambda(p,q,\mu)}{\partial p}=\frac{d}{d p} \left( \frac{S_m}{m} \right)\lambda - \big(s_m + s_n\big)\lambda\epsilon^\lambda\tilde{m}^p \\
&\displaystyle=\left(-\frac{S_m}{m^2}\frac{d m(p)}{d p}+\frac{1}{m}\frac{d S_m(p)}{d p}\right)\lambda - \big(s_m + s_n\big)\lambda\epsilon^\lambda\tilde{m}^p \displaystyle=\left(s_m\tilde{m}^p-1\right)\lambda - \big(s_m + s_n\big)\lambda\epsilon^\lambda\tilde{m}^p \\
&\displaystyle= \left(W_m \tilde{m}^p -\lambda\right) - W\epsilon^\lambda\tilde{m}^p = -\lambda - \tilde{m}^p\left[W_n-W(1-\epsilon^\lambda)\right]
\end{align*}
where \(\displaystyle\frac{d S_m(p)}{d p} = -m(p)\) by the definitions of \(S_m(p)\) and \(m(p)\) in Equation (\ref{equation:def_sm}) and (\ref{equation:m}). Similarly, the impact of the CP-side price \(q\) on social welfare \(W\) is
\[\frac{\partial W(p,q,\mu)}{\partial q} = -\lambda - \tilde{n}^q\left[W_m-W(1-\epsilon^\lambda)\right].\]
When \(\tilde{S}^p_m\) increases with the price \(p\), we have that
\[\frac{d \tilde S^p_m(p)}{d p}=\frac{d}{d p} \left( \frac{m}{S_m} \right)=\frac{\partial}{\partial p} \left( \frac{1}{s_m} \right) = -\frac{d s_m(p)}{d p}\left(\frac{1}{s_m}\right)^2>0\]
and thus \(\displaystyle\frac{d s_m(p)}{d p}<0\) holds.
Based on this condition, we have that
\begin{align*}
\frac{\partial W(p,q,\mu)}{\partial p} = \frac{d s_m(p)}{d p}\lambda + \big(s_m + s_n\big)\frac{\partial \lambda(p,q,\mu)}{\partial p} < \frac{d s_m(p)}{d p}\lambda <0
\end{align*}
implying that \(W\) decreases with \(p\).
Similarly, we can proof that when \(\tilde{S}^q_n\) increases with \(q\), \(W\) decreases with \(q\).
\QEDA\\

\hspace{-0.12in}\textbf{Proof of Theorem \ref{theorem:social-welfare}:} If prices $p$ and $q$ maximize social welfare $W$ and have a fixed total value, we have the first order condition
\[\frac{\partial W(p,q,\mu)}{\partial p} - \frac{\partial W(p,q,\mu)}{\partial q} = 0.\]
By Proposition \ref{proposition:welfare}, we can derive that
\[\frac{\tilde{m}^p}{W_m- W(1 - \epsilon^\lambda)} =\frac{\tilde{n}^q}{W_n- W(1 - \epsilon^\lambda)}.\]
Substituting the relations \(W_m = s_m\lambda\) and \(W_n = s_n\lambda\), we can get the equation that
\[\frac{\tilde{m}^p}{s_m-(s_m+s_n)(1 - \epsilon^\lambda)} =\frac{\tilde{n}^q}{s_n-(s_m+s_n)(1 - \epsilon^\lambda)}\]
and thus Equation (\ref{equation:welfare}) holds.
\QEDA\\

\hspace{-0.12in}\textbf{Proof of Corollary \ref{corollary:profit capacity}:} Under any fixed capacity \(\mu\), the two-sided price \((p,q)\) determines the AP's profit \(U\). Thus the profit is a function of the price and we can denote it by \(U(p,q) \triangleq U(p,q,\mu)\).
We use \(H(p,q)\) to denote the Hessian matrix of the profit function \(U(p,q)\), defined by
\begin{align*}
H(p,q)\triangleq
\begin{bmatrix}
\displaystyle\frac{\partial^2 U(p,q)}{\partial p^2} & \displaystyle\frac{\partial^2 U(p,q)}{\partial p \partial q}\vspace{0.05in}\\
\displaystyle\frac{\partial^2 U(p,q)}{\partial q \partial p} & \displaystyle\frac{\partial^2 U(p,q)}{\partial q^2}
\end{bmatrix}=
\begin{bmatrix}
\displaystyle\frac{\partial^2 U(p,q,\mu)}{\partial p^2} & \displaystyle\frac{\partial^2 U(p,q,\mu)}{\partial p \partial q}\vspace{0.05in}\\
\displaystyle\frac{\partial^2 U(p,q,\mu)}{\partial q \partial p} & \displaystyle\frac{\partial^2 U(p,q,\mu)}{\partial q^2}
\end{bmatrix}.
\end{align*}
We denote the determinant of \(H(p,q)\) by
\[D(p,q) \triangleq det\big(H(p,q)\big) = \frac{\partial^2 U(p,q,\mu)}{\partial p^2}\frac{\partial^2 U(p,q,\mu)}{\partial q^2}-\frac{\partial^2 U(p,q,\mu)}{\partial p \partial q}\frac{\partial^2 U(p,q,\mu)}{\partial q \partial p}.\]
Because the profit-optimal price \((p^*,q^*)\) is unique for any given capacity \(\mu\), the prices \(p^*\) and \(q^*\) are functions of \(\mu\) and thus we can write them as \(p^*(\mu)\) and \(q^*(\mu)\). Moreover, the unique profit-optimal price \((p^*,q^*)\) satisfies the first order condition 
\begin{equation}\label{equation:first pqmu}
\frac{\partial U\big(p^*,q^*,\mu\big)}{\partial p^*} = \frac{\partial U\big(p^*,q^*,\mu\big)}{\partial q^*} = 0,
\end{equation}
thus it is a critical point of the profit function \(U\). Because \((p^*,q^*)\) is also a strict local maximum of \(U\), we have that
\[D(p^*,q^*)= \frac{\partial^2 U(p^*,q^*,\mu)}{\partial (p^*)^2}\frac{\partial^2 U(p^*,q^*,\mu)}{\partial (q^*)^2} - \frac{\partial^2 U(p^*,q^*,\mu)}{\partial q^* \partial p^*}\frac{\partial^2 U(p^*,q^*,\mu)}{\partial p^*\partial q^*}\ge 0\]
by second partial derivative test.

Next, we prove the impact of the capacity \(\mu\) on the optimal prices \(p^*\) and \(q^*\). By the identity in Equation (\ref{equation:first pqmu}),
there exists the identity that
\begin{align*}
\begin{cases}
\displaystyle\frac{d}{d \mu}\left(\frac{\partial U(p^*,q^*,\mu)}{\partial p^*}\right)= \frac{\partial^2 U(p^*,q^*,\mu)}{\partial (p^*)^2}\frac{\partial p^*(\mu)}{\partial \mu} + \frac{\partial^2 U(p^*,q^*,\mu)}{\partial q^* \partial p^*}\frac{\partial q^*(\mu)}{\partial \mu} + \frac{\partial^2 U(p^*,q^*,\mu)}{\partial \mu\partial p^*} = 0\vspace{0.05in}\\
\displaystyle\frac{d}{d \mu}\Big(\frac{\partial U(p^*,q^*,\mu)}{\partial q^*}\Big) = \frac{\partial^2 U(p^*,q^*,\mu)}{\partial p^*\partial q^*}\frac{\partial p^*(\mu)}{\partial \mu} + \frac{\partial^2 U(p^*,q^*,\mu)}{\partial (q^*)^2}\frac{\partial q^*(\mu)}{\partial \mu} + \frac{\partial^2 U(p^*,q^*,\mu)}{\partial \mu\partial q^*} = 0
\end{cases}
\end{align*}
from which we can derive that
\begin{align}\label{equation:basic expression mu}
\begin{cases}
\displaystyle\frac{\partial p^*(\mu)}{\partial  \mu} = \frac{-1}{D(p^*,q^*)}\left(\frac{\partial^2 U(p^*,q^*,\mu)}{\partial (q^*)^2}\frac{\partial^2 U(p^*,q^*,\mu)}{\partial \mu\partial p^*} - \frac{\partial^2 U(p^*,q^*,\mu)}{\partial q^* \partial p^*}\frac{\partial^2 U(p^*,q^*,\mu)}{\partial \mu\partial q^*} \right)\vspace{0.05in}\\
\displaystyle\frac{\partial q^*(\mu)}{\partial  \mu} = \frac{-1}{D(p^*,q^*)}\left(\frac{\partial^2 U(p^*,q^*,\mu)}{\partial (p^*)^2}\frac{\partial^2 U(p^*,q^*,\mu)}{\partial \mu\partial q^*} - \frac{\partial^2 U(p^*,q^*,\mu)}{\partial p^* \partial q^*}\frac{\partial^2 U(p^*,q^*,\mu)}{\partial \mu\partial p^*} \right).
\end{cases}
\end{align}
We denote the inverse function of system congestion \(\varphi\big(m(p),n(q),\mu\big)\) with respect to \(\mu\) by \(\Xi\big(m(p),n(q),\varphi\big)\).
By Equation (\ref{equation:system elasticity}), the elasticity \(\epsilon^\lambda\) of system throughput can be written as a function of \(p,q,\) and \(\varphi\):
\[\epsilon^\lambda(p,q,\varphi) = \left(1-\frac{m(p)n(q)d\rho(\varphi)/d\varphi}{\partial \Lambda(\varphi,\Xi)/\partial \varphi}\right)^{-1}.\]
By Equation (\ref{equation:KKT proof}), we can derive that
\begin{align*}
&\frac{\partial^2 U(p^*,q^*,\mu)}{\partial \mu\partial p^*}  = \frac{\partial}{\partial \mu}\left[\lambda(p^*,q^*,\mu) -(p^*+q^*-c)\epsilon^\lambda(p^*,q^*,\varphi) \lambda(p^*,q^*,\mu) \tilde{m}^p(p^*)\right] \\
&=\frac{\partial \lambda(p^*,q^*,\mu)}{\partial \mu}\left[1\!-\!(p^*\!+\!q^*\!-\!c)\epsilon^\lambda(p^*,q^*,\varphi)\tilde{m}^p(p^*)\right] - (p^*\!+\!q^*\!-\!c)\frac{\partial \epsilon^\lambda(p^*,q^*,\varphi)}{\partial \varphi}\frac{\partial \varphi(p^*,q^*,\mu)}{\partial \mu} \lambda(p^*,q^*,\mu) \tilde{m}^p(p^*)\\
&=  - (p^*+q^*-c)\frac{\partial \epsilon^\lambda(p^*,q^*,\varphi)}{\partial \varphi}\frac{\partial \varphi(p^*,q^*,\mu)}{\partial \mu} \lambda(p^*,q^*,\mu) \tilde{m}^p(p^*).
\end{align*}
Similarly, we can derive that
\begin{align*}
\frac{\partial^2 U(p^*,q^*,\mu)}{\partial \mu\partial q^*} = - (p^*+q^*-c)\frac{\partial \epsilon^\lambda(p^*,q^*,\varphi)}{\partial \varphi}\frac{\partial \varphi(p^*,q^*,\mu)}{\partial \mu} \lambda(p^*,q^*,\mu) \tilde{n}^q(q^*). 
\end{align*}
By Theorem \ref{theorem:KKT-Lerner}, \(\tilde{m}^p(p^*)= \tilde{n}^q(q^*)\) and thus it satisfies that \(\displaystyle\frac{\partial^2 U(p^*,q^*,\mu)}{\partial \mu\partial p^*} = \frac{\partial^2 U(p^*,q^*,\mu)}{\partial \mu\partial q^*}\). Because \(\displaystyle\frac{\partial \varphi(p^*,q^*,\mu)}{\partial \mu}<0\) by Proposition \ref{proposition:elasticity}, we have that
\begin{equation}\label{equation:sign mu}
\sgn\left(\frac{\partial^2 U(p^*,q^*,\mu)}{\partial \mu\partial p^*}\right) = \sgn\left(\frac{\partial^2 U(p^*,q^*,\mu)}{\partial \mu\partial q^*}\right) = \sgn\left(\frac{\partial \epsilon^\lambda(p^*,q^*,\varphi)}{\partial \varphi}\right).
\end{equation}
Besides, by the monotone conditions in Assumption \ref{ass:hazard rate}, we have \(\displaystyle\frac{d \tilde m^p(p^*)}{d p},\displaystyle\frac{d \tilde n^q(q^*)}{d q}>0\), and therefore, by Proposition \ref{proposition:profit}, we can derive that
\begin{align}
\begin{cases}
\displaystyle\frac{\partial^2 U(p^*,q^*,\mu)}{\partial (q^*)^2} - \frac{\partial^2 U(p^*,q^*,\mu)}{\partial q^* \partial p^*} = - (p^*+q^*-c)\lambda(p^*,q^*,\mu)\epsilon^\lambda(p^*,q^*,\varphi)\frac{d \tilde n^q(q^*)}{d q} <0\vspace{0.05in}\\
\displaystyle\frac{\partial^2 U(p^*,q^*,\mu)}{\partial (p^*)^2} - \frac{\partial^2 U(p^*,q^*,\mu)}{\partial p^* \partial q^*} = - (p^*+q^*-c)\lambda(p^*,q^*,\mu)\epsilon^\lambda(p^*,q^*,\varphi)\frac{d \tilde m^p(p^*)}{d p}  < 0.
\end{cases}
\label{equation:hess}
\end{align}
Combining Equation (\ref{equation:basic expression mu}), (\ref{equation:sign mu}) and (\ref{equation:hess}), we can derive that the signs of the marginal prices satisfiy
\begin{align*}
\begin{cases}
\displaystyle\sgn \left(\frac{\partial p^*(\mu)}{\partial \mu} \right)  =  \sgn \left[ \frac{-1}{D(p^*,q^*)}\left(\frac{\partial^2 U(p^*,q^*,\mu)}{\partial (q^*)^2}- \frac{\partial^2 U(p^*,q^*,\mu)}{\partial q^* \partial p^*}\right)\frac{\partial \epsilon^\lambda(p^*,q^*,\varphi)}{\partial \varphi} \right]  =  \sgn\left( \frac{\partial \epsilon^\lambda(p^*,q^*,\varphi)}{\partial \varphi} \right)\vspace{0.05in}\\
\displaystyle\sgn \left(\frac{\partial q^*(\mu)}{\partial \mu} \right)  =  \sgn \left[ \frac{-1}{D(p^*,q^*)}\left(\frac{\partial^2 U(p^*,q^*,\mu)}{\partial (p^*)^2}- \frac{\partial^2 U(p^*,q^*,\mu)}{\partial p^* \partial q^*}\right)\frac{\partial \epsilon^\lambda(p^*,q^*,\varphi)}{\partial \varphi} \right]  =  \sgn\left( \frac{\partial \epsilon^\lambda(p^*,q^*,\varphi)}{\partial \varphi} \right).
\end{cases}
\end{align*}
and the ratio of the marginal prices satisfies
\[\frac{\partial p^*(\mu)}{\partial  \mu} \!:\!  \frac{\partial q^*(\mu)}{\partial \mu} \! =\!  \left(\frac{\partial^2 U(p^*,q^*,\mu)}{\partial (q^*)^2} - \frac{\partial^2 U(p^*,q^*,\mu)}{\partial q^* \partial p^*}\right) \!:\! \left(\frac{\partial^2 U(p^*,q^*,\mu)}{\partial (p^*)^2} - \frac{\partial^2 U(p^*,q^*,\mu)}{\partial p^* \partial q^*}\right)  \!=\!  \frac{d \tilde{n}^q(q^*)}{d q} \!:\! \frac{d \tilde{m}^p(p^*)}{d p}.\tag*{\QEDA}\]

\hspace{-0.12in}\textbf{Proof of Corollary \ref{corollary:welfare capacity}:} Under the constraint \(p+q=c\), the social welfare \(W\) can be denoted by \(W(p,\mu)\triangleq W(p,c-p,\mu) = W(p,q,\mu)\). Because \(p^\circ\) is the unique welfare-optimal price for any given capacity \(\mu\), the price \(p^\circ\) is a function of \(\mu\) and thus we can write it as \(p^\circ(\mu)\). Moreover, the unique profit-optimal price \(p^\circ\) must be a strict local maximum of the welfare function \(W\), thus we have the first order condition for any capacity \(\mu\):
\begin{align}
\frac{\partial W(p^\circ,\mu)}{\partial p^\circ} = &\frac{\partial W(p^\circ,q^\circ,\mu)}{\partial p^\circ} - \frac{\partial W(p^\circ,q^\circ,\mu)}{\partial q^\circ}=\tilde{n}^q(q^\circ)\left[W_m(p^\circ,q^\circ,\mu)-W(p^\circ,q^\circ,\mu)(1-\epsilon^\lambda(p^\circ,q^\circ,\varphi))\right]\notag\\
&-\tilde{m}^p(p^\circ)\left[W_n(p^\circ,q^\circ,\mu)-W(p^\circ,q^\circ,\mu)(1-\epsilon^\lambda(p^\circ,q^\circ,\varphi))\right]=0\label{equation:use}
\end{align}
by Proposition \ref{proposition:welfare}.
So there exists the identity that
\begin{equation*}
\frac{d}{d\mu}\left(\frac{\partial W(p^\circ,\mu)}{\partial p^\circ}\right) = \frac{\partial^2 W(p^\circ,\mu)}{\partial (p^\circ)^2}\frac{\partial p^\circ(\mu)}{\partial \mu} + \frac{\partial^2 W(p^\circ,\mu)}{\partial \mu\partial p^\circ} = 0
\end{equation*}
from we can derive that
\begin{equation}\label{equation:p1}
\frac{\partial p^\circ(\mu)}{\partial \mu} = -\left(\frac{\partial^2 W(p^\circ,\mu)}{\partial (p^\circ)^2}\right)^{-1}\frac{\partial^2 W(p^\circ,\mu)}{\partial \mu\partial p^\circ}. 
\end{equation}
By Equation (\ref{equation:use}), it satisfies that
\begin{align*}
\frac{\partial^2 W(p^\circ,\mu)}{\partial \mu\partial p^\circ} &=\frac{1}{\lambda(p^\circ,q^\circ,\mu)}\frac{\partial \lambda(p^\circ,q^\circ,\mu)}{\partial \mu}\frac{\partial W(p^\circ,\mu)}{\partial p^\circ}-\big(\tilde{m}^p(p^\circ)-\tilde{n}^q(q^\circ)\big)W(p^\circ,q^\circ,\mu)\frac{\partial \epsilon^\lambda(p^\circ,q^\circ,\varphi)}{\partial \varphi}\frac{\partial \varphi(p^\circ,q^\circ,\mu)}{\partial \mu}\\
&=-\big(\tilde{m}^p(p^\circ)-\tilde{n}^q(q^\circ)\big)W(p^\circ,q^\circ,\mu)\frac{\partial \epsilon^\lambda(p^\circ,q^\circ,\varphi)}{\partial \varphi}\frac{\partial \varphi(p^\circ,q^\circ,\mu)}{\partial \mu}.
\end{align*}
Because \(\displaystyle\frac{\partial \varphi(p^\circ,q^\circ,\mu)}{\partial \mu}<0\) from Proposition \ref{proposition:elasticity}, it satisfies that
\[\sgn\left(\frac{\partial^2 W(p^\circ,\mu)}{\partial \mu\partial p^\circ}\right) =  \sgn\big(\tilde{m}^p(p^\circ)-\tilde{n}^q(q^\circ)\big)\cdot\sgn\left(\frac{\partial \epsilon^\lambda(p^\circ,q^\circ,\varphi)}{\partial \varphi}\right).\]
Because \(p^\circ\) is a strict local maximum of \(W\), we have the second order condition \(\displaystyle\frac{\partial^2 W(p^\circ,\mu)}{\partial (p^\circ)^2}<0\). Furthermore, by Equation (\ref{equation:p1}), the sign of the marginal price is
\[\sgn\left(\frac{\partial p^\circ(\mu)}{\partial \mu}\right)= \sgn\left(\frac{\partial^2 W(p^\circ,\mu)}{\partial \mu\partial p^\circ}\right)=  \sgn\big(\tilde{m}^p(p^\circ)-\tilde{n}^q(q^\circ)\big)\cdot\sgn\left(\frac{\partial \epsilon^\lambda(p^\circ,q^\circ,\varphi)}{\partial \varphi}\right).\]
Besides, under the constraint \(p^\circ+q^\circ =c\), we have \(q^\circ(\mu) = c - p^\circ(\mu)\) for any capacity \(\mu\) and thus there exists the identity that
\(\displaystyle\sgn\left(\frac{\partial p^\circ(\mu)}{\partial \mu}\right)= -\sgn\left(\frac{\partial q^\circ(\mu)}{\partial \mu}\right)\).
\QEDA\\

\hspace{-0.12in}\textbf{Proof of Proposition \ref{equation:elasticity relation}:} If the congestion function is \(\Phi(\lambda,\mu)=\lambda/\mu\), its inverse function with respect to \(\lambda\) is \(\Lambda(\phi,\mu) = \phi\mu\). From Equation (\ref{equation:system elasticity}) and Definition \ref{def:elasticity}, the elasticity of system throughput satisfies that
\begin{align*}
\epsilon^\lambda &= \left(1 - \frac{mnd \rho(\varphi)/d\varphi}{\partial \Lambda(\varphi,\mu)/\partial \varphi}\right)^{-1} = \left(1 - \frac{mnd\rho(\varphi)/d \varphi}{\mu}\right)^{-1} \\
&= \left(1 - \frac{mn\varphi}{\lambda}\frac{d \rho(\varphi)}{d \varphi}\right)^{-1}  =   \left(1 - \frac{mn\varphi}{mn\rho}\frac{d \rho(\varphi)}{d \varphi}\right)^{-1}  =  \frac{1}{1 + \epsilon^\rho_\varphi}.\tag*{\QEDA}
\end{align*}

\hspace{-0.12in}\textbf{Proof of Corollary \ref{corollary:profit sensitivity}:} When the gain function is extended to be \(\rho(\phi,s)\), the system congestion and throughput are extended to be \(\varphi(p,q,\mu,s)\) and 
\begin{equation}\label{equation:extension_lambda}
\lambda(p,q,\mu,s)\triangleq \lambda\big(p,q,s,\varphi(p,q,\mu,s)\big) = m(p)n(q)\rho\big(\varphi(p,q,\mu,s),s\big),
\end{equation}
respectively. Correspondingly, we can denote the AP's profit by \(U(p,q,\mu,s) \triangleq (p+q-c)\lambda(p,q,\mu,s)\).
Under any fixed capacity \(\mu\) and sensitivity \(s\), the two-sided price \((p,q)\) determines the AP's profit \(U\). Thus the profit is a function of the price and we can denote it by \(U(p,q) \triangleq U(p,q,\mu,s)\).
We use \(H_s(p,q)\) to denote the Hessian matrix of the profit function \(U(p,q)\), defined by
\begin{align*}
H_s(p,q)\triangleq
\begin{bmatrix}
\displaystyle\frac{\partial^2 U(p,q)}{\partial p^2} & \displaystyle\frac{\partial^2 U(p,q)}{\partial p \partial q}\vspace{0.05in}\\
\displaystyle\frac{\partial^2 U(p,q)}{\partial q \partial p} & \displaystyle\frac{\partial^2 U(p,q)}{\partial q^2}
\end{bmatrix}=
\begin{bmatrix}
\displaystyle\frac{\partial^2 U(p,q,\mu,s)}{\partial p^2} & \displaystyle\frac{\partial^2 U(p,q,\mu,s)}{\partial p \partial q}\vspace{0.05in}\\
\displaystyle\frac{\partial^2 U(p,q,\mu,s)}{\partial q \partial p} & \displaystyle\frac{\partial^2 U(p,q,\mu,s)}{\partial q^2}
\end{bmatrix}
\end{align*}
We denote the determinant of \(H_s(p,q)\) by
\[D_s(p,q) \triangleq det\big(H_s(p,q)\big) = \frac{\partial^2 U(p,q,\mu,s)}{\partial p^2}\frac{\partial^2 U(p,q,\mu,s)}{\partial q^2}-\frac{\partial^2 U(p,q,\mu,s)}{\partial p \partial q}\frac{\partial^2 U(p,q,\mu,s)}{\partial q \partial p}.\]
Because the profit-optimal price \((p^*,q^*)\) is unique for any given capacity \(\mu\) and sensitivity \(s\), the prices \(p^*\) and \(q^*\) are functions of \(\mu\) and \(s\) and thus we can write them as \(p^*(\mu,s)\) and \(q^*(\mu,s)\). Moreover, the unique profit-optimal price \((p^*,q^*)\) satisfies the first order condition 
\begin{equation}\label{equation:first pqmus}
\frac{\partial U\big(p^*,q^*,\mu,s\big)}{\partial p^*} = \frac{\partial U\big(p^*,q^*,\mu,s\big)}{\partial q^*} = 0,
\end{equation}
thus it is a critical point of the profit function \(U\). Because \((p^*,q^*)\) is also a strict local maximum of \(U\), we have that
\[D_s(p^*,q^*)= \frac{\partial^2 U(p^*,q^*,\mu,s)}{\partial (p^*)^2}\frac{\partial^2 U(p^*,q^*,\mu,s)}{\partial (q^*)^2} - \frac{\partial^2 U(p^*,q^*,\mu,s)}{\partial q^* \partial p^*}\frac{\partial^2 U(p^*,q^*,\mu,s)}{\partial p^*\partial q^*}\ge 0\]
by second partial derivative test.
Similar to Equation(\ref{equation:basic expression mu}), we can derive that
\begin{align}\label{equation:basic expression s}
\begin{cases}
\displaystyle\frac{\partial p^*(\mu,s)}{\partial  s} = \frac{-1}{D_s(p^*,q^*)}\left(\frac{\partial^2 U(p^*,q^*,\mu,s)}{\partial (q^*)^2}\frac{\partial^2 U(p^*,q^*,\mu,s)}{\partial s\partial p^*} - \frac{\partial^2 U(p^*,q^*,\mu,s)}{\partial q^* \partial p^*}\frac{\partial^2 U(p^*,q^*,\mu,s)}{\partial s\partial q^*} \right)\vspace{0.05in}\\
\displaystyle\frac{\partial q^*(\mu,s)}{\partial  s} = \frac{-1}{D_s(p^*,q^*)}\left(\frac{\partial^2 U(p^*,q^*,\mu,s)}{\partial (p^*)^2}\frac{\partial^2 U(p^*,q^*,\mu,s)}{\partial s\partial q^*} - \frac{\partial^2 U(p^*,q^*,\mu,s)}{\partial p^* \partial q^*}\frac{\partial^2 U(p^*,q^*,\mu,s)}{\partial s\partial p^*} \right).
\end{cases}
\end{align}
Because the gain function satisfies \(\displaystyle\frac{\partial \rho(\phi,s_1)}{\partial \phi}>\frac{\partial \rho(\phi,s_2)}{\partial \phi}\) for \(\forall s_1<s_2\), i.e., \(\displaystyle\frac{\partial^2 \rho}{\partial\phi\partial s}<0\), we can derive that
\begin{equation}\label{eq:rho_phi_s}
\frac{\partial \rho(\phi,s)}{\partial s} = \int_0^\phi \frac{\partial^2 \rho(t,s)}{\partial t\partial s} dt<0.
\end{equation}
Furthermore, under the identity \(g(p,q,\mu,s)\triangleq \Lambda\big(\varphi(p,q,\mu,s),\mu\big) - \lambda(p,q,\mu,s)=0\), we can derive that
\begin{align*}
&\frac{\partial g(p,q,\mu,s)}{\partial s} = \frac{\partial \Lambda(\varphi,\mu)}{\partial \varphi}\frac{\partial \varphi(p,q,\mu,s)}{\partial s}-\frac{\partial \lambda(p,q,\mu,s)}{\partial s} \\
&= \frac{\partial \Lambda(\varphi,\mu)}{\partial \varphi}\frac{\partial \varphi(p,q,\mu,s)}{\partial s}-\frac{\partial \lambda(p,q,s,\varphi)}{\partial s}-\frac{\partial \lambda(p,q,s,\varphi)}{\partial \varphi}\frac{\partial \varphi(p,q,\mu,s)}{\partial s}=0.
\end{align*}
from which we have
\begin{equation}\label{eq:phi_s}
\frac{\partial \varphi(p,q,\mu,s)}{\partial s} \!=\!  \frac{\partial \lambda(p,q,s,\varphi)}{\partial s} \left(\frac{\partial \Lambda(\varphi,\mu)}{\partial \varphi}\!-\!\frac{\partial \lambda(p,q,s,\varphi)}{\partial \varphi}\right)^{-1}\!\!\! =\!  mn\frac{\partial \rho(\varphi,s)}{\partial s} \left(\frac{\partial \Lambda(\varphi,\mu)}{\partial \varphi}\!-\!\frac{\partial \lambda(p,q,s,\varphi)}{\partial \varphi}\right)^{-1}\!\!< 0.
\end{equation}
Furthermore, by Equation (\ref{equation:system elasticity}), it satisfies that
\begin{equation}\label{eq:epsilon_s}
\frac{\partial \epsilon^\lambda(p,q,\mu,s,\varphi)}{\partial s}= m(p)n(q)(\epsilon^\lambda)^2\frac{\partial^2 \rho(\varphi,s)/\partial \varphi \partial s}{\partial \Lambda(\varphi,\mu)/\partial \varphi}<0.
\end{equation}
Because the inverse \(\Lambda\) of the congestion function satisfies \(\displaystyle\frac{\partial \Lambda(\phi,\mu_1)}{\partial \phi} \le \frac{\partial \Lambda(\phi,\mu_2)}{\partial \phi}\) for all \(\mu_1<\mu_2\), i.e., \(\displaystyle\frac{\partial^2 \Lambda}{\partial\phi\partial\mu}\ge 0\), by Equation (\ref{equation:system elasticity}), we have that
\begin{equation}\label{eq:epsilon_mu}
\begin{aligned}
\frac{\partial \epsilon^\lambda(p,q,\mu,s,\varphi)}{\partial \mu}= -mn(\epsilon^\lambda)^2\frac{\partial \rho(\varphi,s)}{\partial \varphi}\frac{\partial^2\Lambda(\varphi,\mu)}{\partial \varphi\partial \mu}\left(\frac{\partial \Lambda(\varphi,\mu)}{\partial \varphi}\right)^{-2}\ge0.
\end{aligned}
\end{equation}
We denote the inverse of the equilibrium congestion function \(\varphi(p,q,\mu,s)\) with respect to \(\mu\) by \(\Xi(p,q,\varphi,s)\) and denote \(\epsilon^\lambda(p,q,\varphi,s) \triangleq \epsilon^\lambda\big(p,q,\Xi(p,q,\varphi,s),s,\varphi\big)\). By Equation (\ref{eq:epsilon_mu}) and the assumption \(\displaystyle\frac{\partial \epsilon^\lambda(p,q,\varphi,s)}{\partial \varphi}>0\) in Corollary \ref{corollary:profit sensitivity}, we can derive that
\begin{align}
\frac{\partial \epsilon^\lambda(p,q,\mu,s,\varphi)}{\partial \varphi} &=  \frac{\partial \epsilon^\lambda(p,q,\varphi,s)}{\partial \varphi}  -  \frac{\partial \epsilon^\lambda(p,q,\mu,s,\varphi)}{\partial \mu} \frac{\partial \Xi(p,q,\varphi,s)}{\partial \varphi}\notag\\
&>  -  \frac{\partial \epsilon^\lambda(p,q,\mu,s,\varphi)}{\partial \mu} \frac{\partial \Xi(p,q,\varphi,s)}{\partial \varphi} =  -\frac{\partial \epsilon^\lambda(p,q,\mu,s,\varphi)}{\partial \mu}\left(\frac{\partial\varphi(p,q,\mu,s)}{\partial \mu}\right)^{ -1}\ge0.\label{eq:epsilon_phi}
\end{align}
Furthermore, by Equation (\ref{eq:phi_s}), (\ref{eq:epsilon_s}) and (\ref{eq:epsilon_phi}), we derive
\begin{equation}
\frac{\partial \epsilon^\lambda(p,q,\mu,s)}{\partial  s} = \frac{\partial \epsilon^\lambda(p,q,\mu,s,\varphi)}{\partial s} + \frac{\partial \epsilon^\lambda(p,q,\mu,s,\varphi)}{\partial \varphi}\frac{\partial\varphi(p,q,\mu,s)}{\partial s}<0.
\label{equation:congestion_s}
\end{equation}
Based on Proposition \ref{proposition:profit}, Theorem \ref{theorem:KKT-Lerner} and Equation (\ref{equation:congestion_s}), we have the relation that
\begin{align}
\frac{\partial^2 U(p^*,q^*,\mu,s)}{\partial  s\partial p^*} &= -(p^*+q^*-c)\lambda(p^*,q^*,\mu,s)\frac{\partial \epsilon^\lambda(p^*,q^*,\mu,s)}{\partial  s}\tilde{m}^p(p^*)\notag\\
&= -(p^*+q^*-c)\lambda(p^*,q^*,\mu,s)\frac{\partial \epsilon^\lambda(p^*,q^*,\mu,s)}{\partial  s}\tilde{n}^q(q^*) = \frac{\partial^2 U(p^*,q^*,\mu,s)}{\partial  s\partial q^*}>0.\label{equation:pqmus}
\end{align}
Besides, by the monotone conditions in Assumption \ref{ass:hazard rate}, we have \(\displaystyle\frac{d \tilde m^p(p^*)}{d p},\displaystyle\frac{d \tilde n^q(q^*)}{d q}>0\), and therefore, by Proposition \ref{proposition:profit}, we can derive that
\begin{align*}
\begin{cases}
\displaystyle\frac{\partial^2 U(p^*,q^*,\mu,s)}{\partial (q^*)^2} - \frac{\partial^2 U(p^*,q^*,\mu,s)}{\partial q^* \partial p^*} = - (p^*+q^*-c)\lambda(p^*,q^*,\mu,s)\epsilon^\lambda(p^*,q^*,\mu, s)\frac{d \tilde n^q(q^*)}{d q} <0\vspace{0.05in}\\
\displaystyle\frac{\partial^2 U(p^*,q^*,\mu,s)}{\partial (p^*)^2} - \frac{\partial^2 U(p^*,q^*,\mu,s)}{\partial p^* \partial q^*} = - (p^*+q^*-c)\lambda(p^*,q^*,\mu,s)\epsilon^\lambda(p^*,q^*,\mu,s)\frac{d \tilde m^p(p^*)}{d p} < 0.
\end{cases}
\end{align*}
Combining it with Equation (\ref{equation:basic expression s}) and (\ref{equation:pqmus}), we further have
\begin{align*}
\begin{cases}
\displaystyle\frac{\partial p^*(\mu,s)}{\partial  s} = \frac{-1}{D_s(p^*,q^*)}\left(\frac{\partial^2 U(p^*,q^*,\mu,s)}{\partial (q^*)^2} - \frac{\partial^2 U(p^*,q^*,\mu,s)}{\partial q^* \partial p^*}\right)\frac{\partial^2 U(p^*,q^*,\mu,s)}{\partial  s\partial p^*}>0\vspace{0.05in}\\
\displaystyle\frac{\partial q^*(\mu,s)}{\partial  s} = \frac{-1}{D_s(p^*,q^*)}\left(\frac{\partial^2 U(p^*,q^*,\mu,s)}{\partial (p^*)^2} - \frac{\partial^2 U(p^*,q^*,\mu,s)}{\partial p^* \partial q^*}\right)\frac{\partial^2 U(p^*,q^*,\mu,s)}{\partial  s\partial q^*}>0
\end{cases}
\end{align*}
showing that the optimal prices \(p^*\) and \(q^*\) both increase with the congestion sensitivity \(s\) and the ratio of the marginal prices satisfies that
\begin{align*}
\frac{\partial p^*(\mu,s)}{\partial  s} :  \frac{\partial q^*(\mu,s)}{\partial  s}  &=  \left(\frac{\partial^2 U(p^*,q^*,\mu,s)}{\partial (q^*)^2} - \frac{\partial^2 U(p^*,q^*,\mu,s)}{\partial q^* \partial p^*} \right) : \left(\frac{\partial^2 U(p^*,q^*,\mu,s)}{\partial (p^*)^2} - \frac{\partial^2 U(p^*,q^*,\mu,s)}{\partial p^* \partial q^*}\right)\\
 & =  \frac{d \tilde{n}^q(q^*)}{d q} : \frac{d \tilde{m}^p(p^*)}{d p}.\tag*{\QEDA}
 \end{align*}

\hspace{-0.12in}\textbf{Proof of Corollary \ref{corollary:welfare sensitivity}:} When the gain function is extended to be \(\rho(\phi,s)\), we can denote the social welfare by \(W(p,q,\mu,s) \triangleq \big(s_m(p)+s_n(q)\big)\lambda(p,q,\mu,s)\) based on Equation (\ref{equation:extension_lambda}). Under the constraint \(p+q=c\), the social welfare \(W\) can be denoted by \(W(p,\mu,s)\triangleq W(p,c-p,\mu,s) = W(p,q,\mu,s)\). Because \(p^\circ\) is the unique welfare-optimal price for any given capacity \(\mu\) and sensitivity \(s\), the price \(p^\circ\) is a function of \(\mu\) and \(s\) and thus we can write it as \(p^\circ(\mu,s)\). Moreover, the unique profit-optimal price \(p^\circ\) must be a strict local maximum of the welfare function \(W\), thus we have the first order condition for any capacity \(\mu\) and sensitivity \(s\):
\begin{align}
\frac{\partial W(p^\circ,\mu,s)}{\partial p^\circ} \!= &\frac{\partial W(p^\circ,q^\circ,\mu,s)}{\partial p^\circ}\! -\! \frac{\partial W(p^\circ,q^\circ,\mu,s)}{\partial q^\circ}=\tilde{n}^q(q^\circ)\Big[W_m(p^\circ,q^\circ,\mu,s)\!-\!W(p^\circ,q^\circ,\mu,s)\big(1-\epsilon^\lambda(p^\circ,q^\circ,\mu,s)\big)\Big]\notag\\
&-\tilde{m}^p(p^\circ)\Big[W_n(p^\circ,q^\circ,\mu,s)-W(p^\circ,q^\circ,\mu,s)\big(1-\epsilon^\lambda(p^\circ,q^\circ,\mu,s)\big)\Big]=0\label{equation:use2}
\end{align}
by Proposition \ref{proposition:welfare}. Similar to Equation (\ref{equation:p1}), we can derive that
\begin{equation}\label{equation:p2}
\frac{\partial p^\circ(\mu,s)}{\partial  s} = -\left(\frac{\partial^2 W(p^\circ,\mu,s)}{\partial (p^\circ)^2}\right)^{-1} \frac{\partial^2 W(p^\circ,\mu,s)}{\partial  s\partial p^\circ}.
\end{equation}
By Equation (\ref{equation:use2}), we can derive that
\begin{align*}
\frac{\partial^2 W(p^\circ,\mu,s)}{\partial  s\partial p^\circ}& =\frac{1}{\lambda}\frac{\partial \lambda(p^\circ,q^\circ,\mu,s)}{\partial s}\frac{\partial W(p^\circ,\mu,s)}{\partial p^\circ}-\big(\tilde{m}^p(p^\circ)-\tilde{n}^q(q^\circ)\big)W(p^\circ,\mu,s)\frac{\partial \epsilon^\lambda(p^\circ,q^\circ,\mu,s)}{\partial  s}\\
&= -\big(\tilde{m}^p(p^\circ)-\tilde{n}^q(q^\circ)\big)W(p^\circ,\mu,s)\frac{\partial \epsilon^\lambda(p^\circ,q^\circ,\mu,s)}{\partial  s}.
\end{align*}
By Equation (\ref{equation:congestion_s}), it satisfies that
\[\sgn\left(\frac{\partial^2 W(p^\circ,\mu,s)}{\partial  s\partial p^\circ}\right) =  \sgn\big(\tilde{m}^p(p^\circ)-\tilde{n}^q(q^\circ)\big).\]
Because \(p^\circ\) is a strict local maximum of \(W\), we have the second order condition \(\displaystyle\frac{\partial^2 W(p^\circ,\mu,s)}{\partial (p^\circ)^2}<0\). Furthermore, by Equation (\ref{equation:p2}), the sign of the marginal price satisfies that
\[\sgn\left(\frac{\partial p^\circ(\mu,s)}{\partial  s}\right)= \sgn\left(\frac{\partial^2 W(p^\circ,\mu,s)}{\partial  s\partial p^\circ}\right) = \sgn\big(\tilde{m}^p(p^\circ)-\tilde{n}^q(q^\circ)\big).\]
Besides, under the constraint \(p^\circ+q^\circ =c\), we have \(q^\circ(\mu,s) = c - p^\circ(\mu,s)\) for any \(\mu,s\) and thus there exists the identity that
\(\displaystyle\sgn\left(\frac{\partial p^\circ(\mu,s)}{\partial s}\right)= -\sgn\left(\frac{\partial q^\circ(\mu,s)}{\partial s}\right)\).
\QEDA

\bibliographystyle{ACM-Reference-Format}
\bibliography{ref}

\end{document}